\documentclass{aa}  
 
\usepackage{graphicx}
 
\usepackage{amsmath}
\usepackage{amssymb}
\usepackage{txfonts,captcont,subfigure}
\usepackage{alltt}
 
\usepackage{natbib} 
\bibpunct{(}{)}{;}{a}{,}{,}
\usepackage[below]{placeins}

\usepackage{url}

\def\etal{{\it et al}.}

\def\hmpc{h^{-1} {\rm Mpc}}

\def\textindent#1{\indent{#1\enspace}\ignorespaces}
\def\itemitem{\par\indent \hangindent2\parindent \textindent}

\newcommand{\cgal}{\texttt{CGAL}\ }
\def\hmpc{~h^{-1} {\rm Mpc}}

\begin{document}
 
  \title{The Multiscale Morphology Filter:\\
    Identifying and Extracting\\Spatial Patterns in the Galaxy Distribution}
 
   \author{Miguel A. Arag\'on-Calvo, Bernard J.T. Jones, Rien van de Weygaert, \and J.M.(Thijs)~van~der~Hulst\inst{1}}
 
   \offprints{M. Arag\'on-Calvo, \email{miguel@astro.rug.nl}}
 
   \institute{Kapteyn Astronomical Institute, University of Groningen,
     P.O. Box 800, 9700 AV Groningen, The Netherlands\\}
 
   \date{Received ...; accepted ...}
 
 
  \abstract
   {} 
   {We present here a new method, MMF, for automatically segmenting cosmic structure into its basic 
    components: clusters, filaments, and walls.  Importantly, the segmentation is scale independent, 
    so all structures are identified without prejudice as to their size or shape.
    The method is ideally suited for extracting catalogues of clusters, walls, and 
    filaments from samples of galaxies in redshift surveys or from particles 
    in cosmological N-body simulations: it makes no prior assumptions about the scale or shape
    of the structures.} 
   {Our Multiscale Morphology Filter (MMF) method has been developed on the basis 
    of visualization and feature extraction techniques in computer vision and medical
    research. The density or intensity field of the sample is smoothed over a range of 
    scales. The smoothed signals are processed through a morphology response 
    filter whose form is dictated by the particular morphological feature it seeks to extract, 
    and depends on the local shape and spatial coherence of the intensity field. The morphology 
    signal at each location is then defined to be the one with the maximum response across 
    the full range of smoothing scales. 

    The success of our method in identifying 
    anisotropic features such as filaments and walls depends critically on the use of an optimally     
    defined intensity field.  This is accomplished by applying the DTFE reconstruction methodology
    to the sample particle or galaxy distribution.}
   {We have tested our MMF Filter against a set of heuristic models of weblike patterns such as are seen in the 
    Megaparsec cosmic matter distribution.  To test its effectiveness in the context of more realistic 
    configurations we also present preliminary results from the MMF analysis of an N-body model.  
    Comparison with alternative prescriptions for feature extraction shows that MMF is a remarkably
    strong structure finder}
   {}
 
 
\titlerunning{The Multiscale Morphology Filter}
\authorrunning{Arag\'on et al.}
\maketitle
%

%
%
\section{Introduction}
On scales from a few Megaparsecs up to more than a hundred Megaparsecs, the spatial cosmic matter distribution displays a salient and pervasive weblike pattern which is perceived in the first instance as a cellular structure. The distribution of galaxies in large scale redshift surveys such as the 2-deg Field Galaxy Redshift Survey \citep[2dF:][]{colless2003} and the Sloan Digital Sky Survey \citep[SDSS:][]{york2000} clearly delineate this \textit{Cosmic Web} \citep{bondweb1996} (see \citet{weygaert2002} for a review).

Large computer simulations of the evolution of cosmic structure \citep{springmillen2005} show prominent cellular patterns arising from gravitational instability.  Galaxies accumulate in flattened walls, elongated filaments and dense compact clusters.  These structures surround large near-empty void regions \citep{zeldovich1982}. Their spatial distribution displays a distinctive frothy texture, interconnected in a cosmic weblike pattern.

While it is rather straightforward to find qualitative descriptions of the spatial structure and components of the cosmic web, a useful, and physically meaningful, quantitative analysis has proven to be far from trivial.  This would be important, for example, when we wish to study the effect of environment on the formation of galaxies and their halos.  
 
\subsection{Multi-scale analysis}
We present here a new method for automatically segmenting cosmic structure into its basic components: clusters, filaments, and walls.  Importantly, the segmentation is scale independent, so all structures are identified without prejudice as to their size or shape.
 
There are two parts to this: firstly, the reconstruction of a continuous density field from a point sample and secondly, the identification of structures within that density field.  For the first part we use the Delaunay Tessellation Field Estimator (\textit{DTFE}) technique of \citet{schaapwey2000}.  The second part, which is the main thrust of this paper, consists of a series of morphology filters that identify, in a scale independent manner, particular kinds of structure in data.  The method is referred to as the Multiscale Morphology Filter (\textit{MMF}) and is based on the kind of Scale Space analysis that has in recent years proved so successful in imaging science.  
 
It is worth emphasising at this juncture that we have chosen a specific implementation of this kind of multi-scale analysis.  Our choice is made on the following grounds: (a) it is simple to understand and program, (b) it works under quite general conditions and (c) the approach is generic and easy to modify.  There are many alternative multi-scale strategies: we leave those for another day or for other people to follow up.  Thus we shall try to keep this presentation as general as possible so that the points at which we make implementation specific choices are clear.

\subsection{Emergence of hierarchical web-like structure}
Structure in the Universe emerged as a result of the gravitational growth of small amplitude primordial density and velocity perturbations. Following the initial linear growth of the Gaussian primordial perturbations, the gravitational clustering process leads to the emergence of complex patterns and structures in the density field. At least three characteristics of the midly nonlinear cosmic matter stand out. 
 
The most prominent property is its hierarchical nature. The gravitational clustering process proceeds such that small structures are the first to materialize and subsequently merge into ever larger entities. As a result each emerging cosmic structure consists of various levels of substructure. Hence, upon seeking to identify structure at one characteristic spatial scale we need to take into account a range of scales. 
 
The second prominent aspect is that of the \textit{weblike geometry} marked by highly elongated filamentary and flattened planar structures.  The existence of the cosmic web can be understood through the tendency of matter concentrations to contract and collapse gravitationally in an anisotropic manner. 
 
A final conspicuous aspect is that of the dominant presence of large roundish underdense regions, the \textit{voids}. They form in and around density troughs in the primordial density field. 
 
The challenge for any viable analysis tool is to trace, highlight and measure these features of the cosmic web. 
 
\subsection{Outline of this paper}
We start in section~\ref{sec:DTFE} by reviewing the DTFE method that is used to sample discrete point sets onto a regular mesh.  Then in section~\ref{sec:ScaleSpace} we introduce the basic ideas from scale space theory that we will use. In section \ref{sec:ScaleSpace} we introduce the morphology filters and give them a geometrical interpretation. The filters are tested using a Voronoi model in section \ref{sec:Voronoi}.  We present 
brief results from an N-body simulation in section~\ref{sec:nbody}, leaving a detailed study to a subsequent paper in this series.  
 
%
%
\section{Structure finding}
Many attempts to describe, let alone identify, the features and components of the Cosmic Web have been of a mainly heuristic nature. 
There is a variety of statistical measures characterizing specific aspects of the large scale matter distribution \citep[for an 
extensive review see][]{martinez2002}. For completeness and comparison, we list briefly a selection of methods for structure characterisation 
and finding.  It is perhaps interesting to note two things about this list: 
\begin{enumerate}
\item[a)] each of the methods tends to be specific to one particular structural entity 
\item[b)] there are no explicit wall-finders. 
\end{enumerate}
This emphasises an important aspect of our Scale Space approach: it provides a uniform approach to finding Blobs, Filaments and Walls 
as individual objects that can be catalogued and studied.
 
\subsection{Structure from higher moments}
The clustering of galaxies and matter is most commonly described in terms of a hierarchy of correlation functions. The two-point correlation function (and its Fourier transform, the power spectrum) remains the mainstay of cosmological clustering analysis and has a solid physical basis.  However, the nontrivial and nonlinear patterns of the cosmic web are mostly a result of the phase correlations in the cosmic matter distribution \citep{rydengram1991,chiang2000,pcoles2000}.  While this information is contained in the moments of cell counts \citep{peebles1980,lapparent1991,gaztanaga1992} and, more formally so, in the full hierarchy of M-point correlation functions $\xi_M$, their measurement has proven to be impractical for all but the lowest orders  \citep{peebles1980,szapudi1998,jones2005}.
 
The Void probability Function \citep{white1979,lachieze1992} provided a characterisation the ''voidness'' of the Universe in terms of a function that combined information from many higher moments of the point distribution.  But, again, this has not provided any identification of individual voids.
 
\subsection{Topological methods}
The shape of the local matter distribution may be traced on the basis of an analysis of the statistical properties of its inertial moments 
\citep{babul1992,vishniac1995,basilakos2001}. These concepts are closely related to the full characterization of the topology of the matter distribution in terms of four Minkowski functionals \citep{mecke1994,schmalzing1999}. They are solidly based on the theory of spatial statistics and also have the great advantage of being known analytically in the case of Gaussian random fields. In particular, the \textit{genus} of the density field has received substantial attention as a strongly discriminating factor between intrinsically different spatial patterns \citep{gott1986,hoyle2002}.

The Minkowski functionals provide global characterisations of structure. An attempt to extend its scope towards providing locally defined topological measures of the density field has been developed in the SURFGEN project defined by Sahni and Shandarin and their coworkers \citep{sahni1998,shandarin2004}. The main problem remains the user-defined, and thus potentially biased, nature of the continuous density field inferred from the sample of discrete objects. The usual filtering techniques suppress substructure on a scale smaller than the filter radius, introduce artificial topological features in sparsely sampled regions and diminish the flattened or elongated morphology of the spatial patterns. Quite possibly the introduction of more advanced geometry based methods to trace the density field may prove a major advance towards solving this problem.
 
Importantly, \citet{martinez2005} and \citet{saar2007} have generalized the use of Minkowski Functionals by calculating their values in a hierarchy of scales generated from wavelet-smoothed volume limited subsamples of the 2dF catalogue. This approach is particularly effective in dealing with non-Gaussian point distributions since the smoothing is not predicated on the use of Gaussian smoothing kernels.
 
\subsection{Cluster finding}
In the context of analysing distributions of galaxies we can think of cluster finding algorithms.  There we might define a cluster as an aggregate of neighbouring galaxies sharing some localised part of velocity space.  Algorithms like HOP attempt to do this.  However, there are always issues arising such as how to deal with substructure: that perhaps comes down to the defintion of what a cluster is. Here we focus on defining coherent structures based on particle positions alone.  The velocity space data is not used since there is no prior prejudice as to what the velocity space should look like. 
\subsection{Filament finding}
The connectedness of elongated supercluster structures in the cosmic matter distribution was first probed by means of percolation analysis, introduced and emphasized by Zel'dovich and coworkers \citep{zeldovich1982}, while a related graph-theoretical construct, the minimum spanning tree of the galaxy distribution, was extensively probed and analysed by Bhavsar and collaborators \citep{barrow1985,graham1995,colberg2007} in an attempt to develop an objective measure of filamentarity. 
 
Finding filaments joining neighbouring clusters has been tackled, using quite different techniques, by \citet{colberg2005} and by \citet{pimbblet2005}.  More general filament finders have been put forward by a number of authors.  Skeleton analysis of the density field \citep{novikov2006} describes continuous density fields by relating density field gradients to density maxima and saddle points. This is computationally intensive but quite effective, though it does depend on the artefacts in the reconstruction of the continuous density field.
 
\citet{stoica2005} use a generalization of the classical Candy model to locate and catalogue filaments in galaxy surveys.  This approach has the advantage that it works directly with the original point process and does not require the creation of a continuous density field.  However, it is very computationally intensive.

\subsection{Void Finding}
Voids are distinctive and striking features of the cosmic web, yet finding them systematically in surveys and simulations has proved rather difficult.  There have been extensive searches for voids in galaxy catalogues \citep{hoyle2002,plionis2002} and in numerical simulations 
\citep{arbmul2002,aikmah1998}.
 
Several factors contribute to making systematic void-finding difficult.  The fact that voids are almost empty of galaxies means that the sampling density plays a key role in determining what is or is not a void \citep{schmidt2001}.  Moreover, void finders are often predicated on building void structures out of cubic cells \citep{kauffair1991} or out of spheres \citep[e.g:][]{patiri2006}.  Such methods attempt to synthesize voids from the intersection of cubic or spherical elements and do so with varying degrees of success.  The \textit{Aspen-Amsterdam Void Finder Comparison Project} of \citet{colpear2007} will clarify many of these issues.
 
The Watershed-based algorithm of \citet{platen2007} aims to avoid issues of both sampling density and shape.

\subsection{Structure from Scale Space}
Combining the local Hessian matrix eigenvalues on various scales, this is the new technique that we are presenting here for the first time in the cosmological context.
 
Scale space analysis looks for structures of a mathematically specified type in a hierarchical, scale independent, manner.  It is presumed that the specific structural characteristic is quantified by some appropriate parameter (e.g.: density, eccentricity, direction, curvature components).  The data is filtered to produce a hierarchy of maps having different resolutions, and at each point, the dominant parameter value is selected from the hierarchy to construct the scale independent map.  We refer to this scale-filtering processes as a \textit{Multiscale morphology filter}.  
 
For simplicity, the paper describes one specific implementation, or embodiment, of the process in relation to the problem of cataloguing the structural elements of the cosmic web.  Other embodiments are possible, but the present one turns out to be highly effective in structure segregation and feature identification.
 
While this sounds relatively straightforward, in practise a number of things are required to execute the process.  Firstly there must be an unambiguous defintion of the structure-defining characteristic.  In the present case we shall use the principal components of the local curvature of the density field at each point as a morphology type indicator.  This requires that the density be defined at all points of a grid, and so there must be a method for going from a discrete point set to a grid sampled continuous density field.  We choose to do this using the DTFE methodology since that does minimal damage to the structural morphology of the density field.
 
Since we are looking for three distinct structural morphologies, blobs, walls and filaments, we have to apply the segmentation process three times.  However, since we shall be using curvature components as structural indicators, we shall have to eliminate the blobs before looking for filaments, and we shall then have to eliminate the filaments before looking for walls.

%
%
%
\begin{figure*}[t]
  \centering
    \includegraphics[width=0.48\textwidth,angle=0.0]{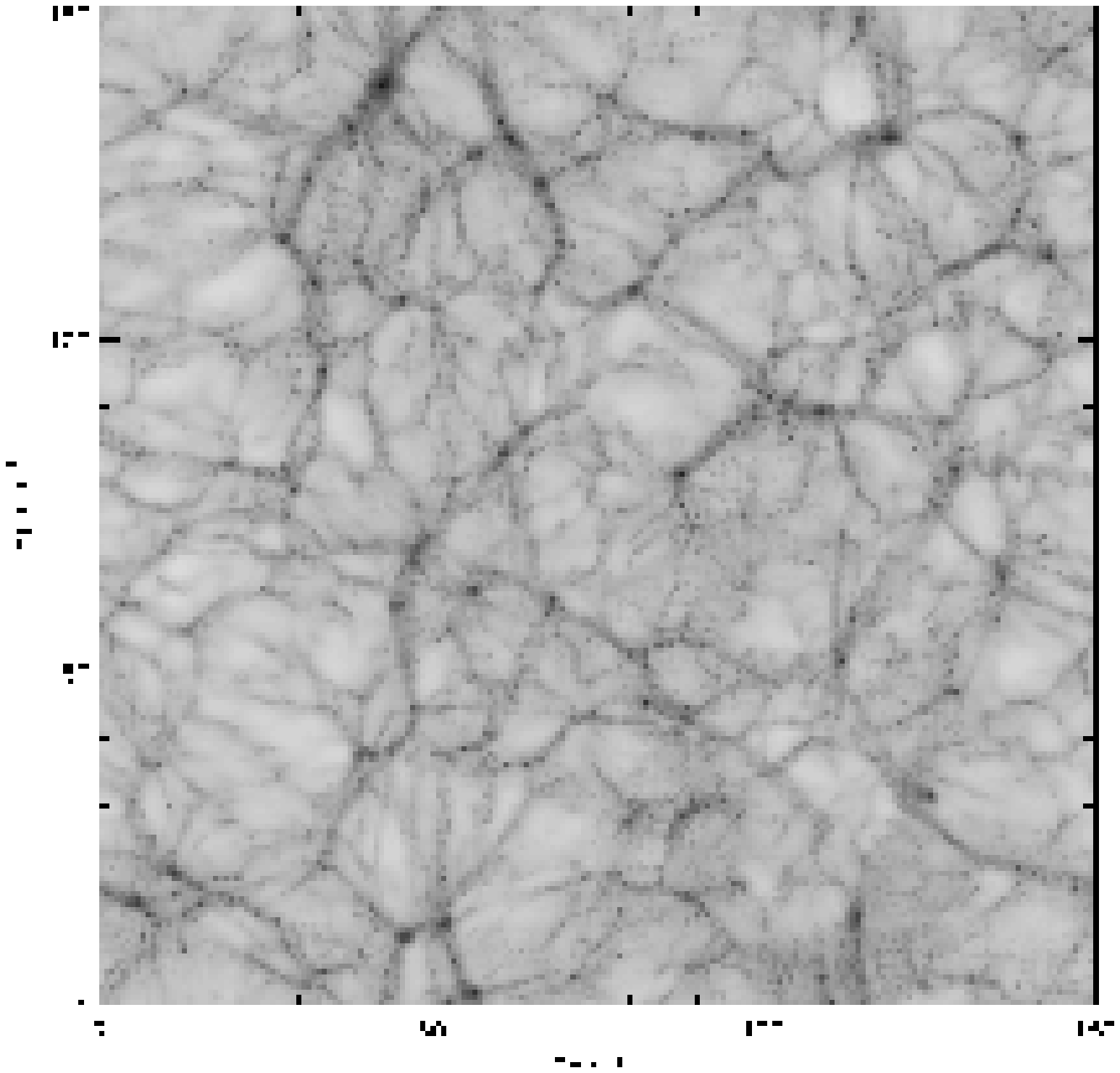}
    \includegraphics[width=0.48\textwidth,angle=0.0]{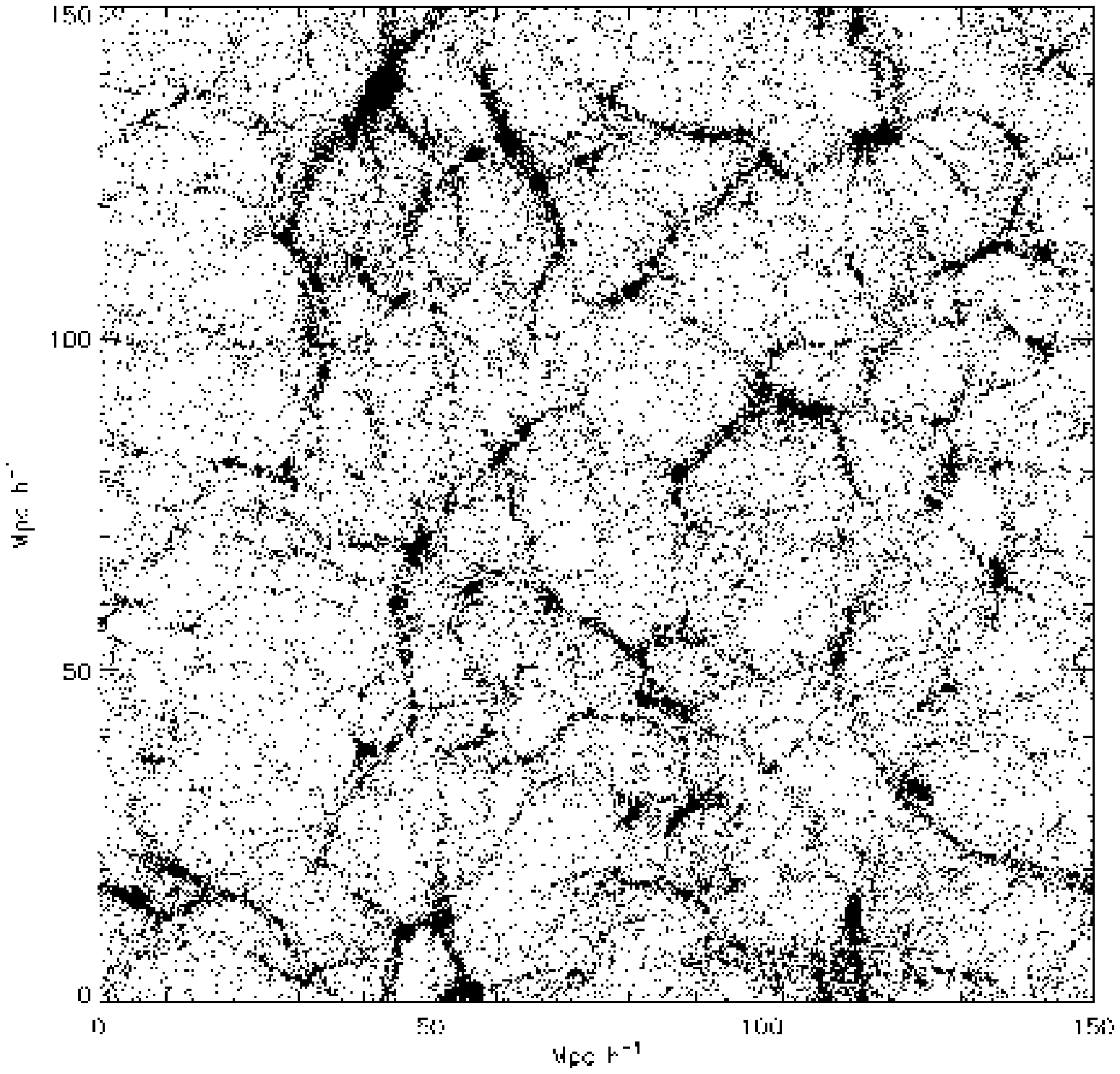}
    \caption{DTFE image of a slice through the N-body simulation used in this work. Left: DTFE density field in a central slice. Right: the corresponding particle distribution in a slice of width $5h^{-1}\hbox{Mpc}$.}
  \label{fig:dtfe} 
\end{figure*} 
\section{Resampling and Rescaling Point sets}
\label{sec:DTFE}
The cosmological problem presents its own difficulties, not the least of which is the fact that the data set is presented, not as a density field, but as a set of discrete points which are presumed to sample some underlying density field.  However, the filtering procedures we use here for defining objects act on continuous fields (or images) and require continuous first and second derivatives of field values.  It is therefore necessary to resample the point set data on a grid.  In doing this we need to assure ourselves that the objects, structures, features and patterns in these fields are resampled in an optimal way: both substructure and morphological characteristics must be preserved.  To achieve this we use the DTFE reconstruction of the density field.

\subsection{The DTFE density field}
The Delaunay Triangulation Field Estimator (``DTFE'') \citep{schaapwey2000,willemphd2007} is a powerful new method, based upon concepts from computational geometry  \citep{okabe2000} that offers a ``safe'' reconstruction in that it accurately preserves the local features.  DTFE produces a morphologically unbiased and optimized continuous density field retaining all features visible in a discrete galaxy or particle distribution. 
 
The input samples for our analysis are mostly samples of galaxy positions obtained by galaxy redshift surveys or the positions of a large number of particles produced by N-body simulations of cosmic structure formation. In order to define a proper continuous field from  discrete distribution of points - computer particles or galaxies - we translate the spatial point sample into a continuous density field by means of the Delaunay Tessellation Field Estimator \cite{schaapwey2000,willemphd2007}. 
 
The DTFE technique recovers fully volume-covering and volume-weighted continuous fields from a discrete set of sample field values. The method has been developed by \citep[][,see]{schaapwey2000,willemphd2007} and forms an elaboration of the velocity interpolation scheme introduced by \cite{bernwey1996}. It is based on the use of the Voronoi and Delaunay tessellations of a given spatial point distribution.  It provides a basis for a natural, fully self-adaptive filter in which the Delaunay tessellations are used as multidimensional interpolation intervals. 
 
The primary ingredient of the DTFE method is the Delaunay tessellation of the particle distribution.  The Delaunay tessellation of a point set is the uniquely defined and volume-covering tessellation of mutually disjunct Delaunay tetrahedra. A Delaunay tetrahedron is defined by the set of four points whose circumscribing sphere does not contain any of the other points in the generating set \citep{delaunay1934} (triangles in 2D). The Delaunay tessellation is intimately related to the Voronoi tessellation of the point set, they are each others {\it dual}. The Voronoi tessellation of a point set is the division of space into mutually disjunct polyhedra, each polyhedron consisting of the part of space closer to the defining point than any of the other points \citep{voronoi1908,okabe2000}.
 
DTFE exploits three particular properties of Voronoi and Delaunay tessellations. The tessellations are very sensitive to the local point density. The DTFE method uses this fact to define a local estimate of the density on the basis of the inverse of the volume of the tessellation cells. Equally important is their sensitivity to the local geometry of the point distribution, which allows them to trace anisotropic features such as encountered in the cosmic web. Finally it uses the adaptive and minimum triangulation properties of Delaunay tessellations to use them as adaptive spatial interpolation intervals for irregular point distributions.  In this it is the first order version of the {\it Natural Neighbour method} \citep[NN method:][]{sibson1980,sibson1981,watson1992, braunsambridge1995, sukumarphd1998, okabe2000}. 
 
One of the important - and crucial - properties of a processed DTFE density field is that it is capable of delineating {\it three} fundamental characteristics of the spatial structure of the Megaparsec cosmic matter distribution. It outlines the {\it full hierarchy of substructures} present in the sampling point distribution, relating to the standard view of structure in the Universe having arisen through the gradual hierarchical buildup of matter concentrations. DTFE also reproduces any {\it anisotropic patterns} in the density distribution without diluting their intrinsic geometrical properties. This is a great advantage when seeking to analyze the cosmic matter distribution, characterized by prominent filamentary and wall-like components 
linking up into a {\it cosmic web}. A third important aspect of DTFE is that it outlines the presence and shape of voidlike regions. DTFE renders the low-density regions as regions of slowly varying, moderately low density values through the interpolation definition of the DTFE field reconstruction. 
 
An outline of the DTFE reconstruction procedure can be found in appendix~\ref{app:dtfe_recons}.
 
\subsection{Rescaling}
In building the scale space we need to construct a hierarchy of rescaled replicas of the original grid-sampled data.  In this paper this is done simply by applying a hierarchy of isotropic Derivative of Gaussian smoothing filters to the data.
 
Of course, substructure and morphological characteristics will be altered during this hierarchical smoothing process.  The smearing of features through smoothing is inevitable if we smooth using isotropic filters and there has been some discussion as to whether one might do better by rescaling in such a way as to minimise feature smearing \citep[for example ][]{martinez2005,saar2007}.  It is possible to use refined (nonlinear) smoothing procedures that minimize the side effects of smoothing but that issue is not addressed here.  Here, we simply rescale using isotropic Gaussian filters: this seems to work very well and avoids complications arising from using other filters.

%
%
\section{Scale Space Analysis}
\label{sec:ScaleSpace}
In this contribution we introduce a method for recognizing and identifying features in data based on the use of a ''Scale Space'' representation of the data \citep{florack1992,lindeberg1998}.  The Scale Space representation of a data set consists simply of a sequence of copies of the data having different resolutions.  A feature searching algorithm is applied to all of these copies, and the features are extracted in a scale independent manner by suitably combining the information from all copies.
 
We use a particular feature recognition process based on eigenvalues of the Hessian matrix of the density field.  It should be understood that the technique we describe here could well be used with other feature recognition systems, such as, for example, the ShapeFinder process \citep{sahni1998}.  Scale Space is a powerful tool for scale independent data analysis.
 
\subsection{Image processing}
The use of this technique can be traced back to the work of David Marr at M.I.T in the 1970's \citep{marrhild1980}, reviewed in his seminal book on the physiology of image understanding: \textit{Vision} \citep{marr1980}. There (\textit{loc. cit.} Chapter 2, especially figures 2-10 and 2-23) he describes what is called the ``Primal Sketch" and the use of what today are called ``Marr Wavelets'' in extracting scale independent information.  We apply precisely this transformation to a scale space representation of a cosmological density field, and in doing so ostensibly extract features in much the same way, according to Marr, that the human visual cortex does.
 
More recently, \cite{frangi1998} and \cite{sato1998} used Scale Space analysis for detecting the web of blood vessels in a medical image.  The vascular system is a notoriously complex pattern of elongated tenuous features whose branching make it closely resemble a fractal network.  We translate, extend and optimize this technology towards the recognition of the major characteristic structural elements in the Megaparsec matter distribution.  The resulting methodology yields a unique framework for the combined identification of dense, compact bloblike clusters, of the salient and moderately dense elongated filaments and of tenuous planar walls.

\subsection{Multiscale Structure Identification}
\label{sec:Identification}
Segmentation of a density field into distinct, meaningful, components has been one of the major goals of 
image processing over the past decades.  There are two stages involved: firstly providing a criterion describing 
the basis for the segmentation, be it colour, texture, motion or some other attribute and secondly providing 
an algorithm whereby those distinguishing attributes can be automatically and unambiguously identified.  
Ambiguities in structure finding frequently occur when the sought-for structure exists on a variety of 
scales that may be nested hierarchically.  
\begin{table*}[t]
\begin{center}
   Maps used in morphological analysis \\
   \begin{tabular}[t]{|c|l|l|l|}
   \hline
   Symbol        & Name                    & Description                                   & Eqn \\
   \hline
                 &                         &                                               & \\
   $\Phi$        & Scale Space Map         & Combination filtered density maps $f_{\rm S,n}$ over all levels $n$. & (\ref{eq:scalespace})\\ 
                 &                         &                                               & \\
                 &                         &                                               & \\
 
   $\mathcal{E}$ & Morphology Mask         & Region of space obeying shape constraint. & (\ref{eq:morphmask})\\ 
                 &                         & E=1: locations obeying shape constraint   & (\ref{tab:morphmask})\\
                 &                         & E=0: locations not obeying shape constraints & \\
 
   $\mathcal{S}$ & Shape Significance Map  & Feature shape fidelity for each point locale. & \\ 
                 &                         & Measures conformance to local shape criteria  & (\ref{eq:significance})\\  
                 &                         &                                               & \\
   $\mathcal{M}$ & Morphology Map          & Soft thresholded version of $\mathcal{S}$.  The threshold selects out the most & \\
                 &                         & locally shape conformant features.  Requires input of a threshold parameter $\beta$ & (\ref{eq:G_geom_1}) \\
                 &                         &                                               & \\
   $\mathcal{I}$ & Morphology Intensity Map   & Map of $\lambda_3$ for blobs,     
                                             $\lambda_2$ for filaments or 
                                             $\lambda_1$ for walls                         & \\
                 &                         & Modulates Morphology map, meant to avoid enhancing noisy low intensity structures 
                                             & (\ref{eq:intensity}) \\
                 &                         &                                               & \\
 
   $\mathcal{T}$ & Morphology Filter       & Constructed from $\mathcal{I}$ and $\mathcal{M}$. Morphology weighted filter & \\
                 &                         & for the Morphology Mask. Provides each location which obeys the morphology constraint  & \\
                 &                         & with a measure of the strength of morphology signal. & (\ref{eq:morphfilter}) \\
                 &                         &                                               & \\
   $\mathcal{F}$ & Feature Map             & Product of morphology mask $\mathcal{E}$ and corresponding morphology filter $\mathcal{T}$.  & \\
                 &                         & There is one Feature Map for each level in the Scale-Space, representing local structures as & \\
                 &                         & seen on the different scales of the Scale-Space & (\ref{eq:featuremap}) \\
                 &                         &                                               & \\
   $\Psi$        & Scale-Space Map Stack   & Constructed from the $\mathcal{F}_i$ for all levels in the Scale-Space   &   \\
                 &                         & Each pixel in this map is the greatest value of the corresponding pixels & \\
                 &                         &  in  the Feature maps that make up the Scale-Space stack & (\ref{eq:featurestack}) \\
                 &                         &                                               & \\
                 &                         &                                               & \\
   $\mathcal{O}$ & Object Map              & Inclusion of astrophysical \& cosmological criteria to select physically recognizable objects & \\
                 &                         & Produced by thresholding Scale-Space Map Stack $\Psi$ &   \\
                 &                         & Threshold criterion determined by cosmological/astrophysical considerations & 
(sec.~\ref{sec:FeatureDetection}) \\
                 &                         &                                               & \\
  \hline
   \end{tabular}
 
\end{center} 
\end{table*}
 
\subsection{the Multiscale Morphology Filter: Outline}
The technique presented here, the Multiscale Morphology Filter (MMF), looks to synthesize global structures by identifying local structures on a variety of scales and assembling them into a single scale independent structural map.  The assembly is done by looking at each point and asking which of the structures found on the various search scales dominates the local environment. This is the essence of the so-called {\it Scale Space approach}. We first provide an outline of the various stages involved with the MMF method. In the subsequent sections we treat various aspects in more detail. 
  
\subsection{The Analysis Cycle}
We are looking for three distinct morphologies within the same distribution.  This requires three passes through the data, each time eliminating the features found in the previous pass.  In the first pass, the blobs in the dataset are identified along with their enclosed datapoints.  The points that are in blobs are eliminated and then the filaments are identified with their constituent points.  After eliminating the filament points the walls and their constituent points can be identified.
 
\noindent
Each pass involves the following components and procedures:
\begin{enumerate}
\item[$\bullet$] {\it Point Dataset}\\ 
For each pass this is the set of galaxies or particles in an N-body model from which we are going to extract a specified feature.  In the first pass this is the full data sample within which we are going to identify blobs.  On the second pass it is the original point set from which the points in the blobs have been removed.  Likewise for the third pass.\\
 
\item[$\bullet$] {\it DTFE Density Field}\\ 
The discrete spatial distribution of galaxies, or particles in a N-body computer model, is resampled to give a continuous volume-filling density field map $f_{\tiny{\textrm{DTFE}}}$ on a high resolution grid. In order to guarantee an optimal representation of morphological features this is accomplished on the basis of the DTFE method \citep{schaapwey2000,willemphd2007}.\\
 
\item[$\bullet$] {\it Scale filtering}\\ 
The raw DTFE density field $f_{\tiny{\textrm{DTFE}}}$ is filtered over a range of spatial scales $R_n$ in order to produce a family $\Phi$ of smoothed density maps $f_S,n$, each defining a level of the {\it Scale-Space} representation. The range of scales is set by the particular interest of the analysis.\\ 
 
\item[$\bullet$] {\it Hessian \& Eigenvalues}\\ 
The Hessian matrix $\nabla_{ij} f_S$ of the density field is computed at each point of each of the smoothed density fields in the filtered Scale-Space density maps $f_S$.  At each point the eigenvalues $\lambda_{k}$ ($k=1,2,3$) of the Hessian matrix are determined.\\ 
 
\item[$\bullet$] {\it Morphology Mask}\\ 
The {\it Morphology Mask} ${\mathcal E}_{\tiny{\textrm{morph}}}$  identifies the locations obeying the required morphology/shape constraints. At every location in every map, ${\mathcal E}=1$ if the shape constraint is valid, ${\mathcal E}=0$ if it does not. This is a ``hard'' filter.\\
 
\item[$\bullet$] {\it Shape Significance Map} \\ 
A {\it Feature shape Significance (or fidelity)} index ${\mathcal S}_{\tiny{\textrm{morph}}}$ is determined for the specified morphology.  This is done on the basis of the signs and ratios of the three eigenvalues $\lambda_{k}$ ($k=1,2,3$), and thus dependent only on the local variations of the field on the various scales present in the scale space maps\\
 
\item[$\bullet$] {\it Morphology Response Map} \\ 
The {\it Morphology Response Filter}, ${\mathcal M}_{\tiny{\textrm{morph}}}$, is the soft thresholded version of the shape significance map ${\mathcal S}_{\tiny{\textrm{morph}}}$.  It selects out the most locally shape conformant features and is computed for each scale space level by processing ${\mathcal S}_{\tiny{\textrm{morph}}}$, weighted by a specified threshold parameter $\beta$.\\ 
 
\item[$\bullet$] {\it Morphology Intensity Map} \\ 
In order to avoid enhancing noisy low intensity structures we include a {\it Morphology Intensity function} $\mathcal{I}_{\tiny{\textrm{morph}}}$ that modulates the morphology response map according to some measure of the feature strength.  We characterise feature strength by the values of the specifc eigenvalues: $\lambda_1$ for the walls, $\lambda_2$ for the filaments and $\lambda_3$ for the blobs.\\
 
\item[$\bullet$] {\it Morphology Filter} \\ 
Morphology weighted filter ${\mathcal T}_{\tiny{\textrm{morph}}}$ for the Morphology Mask ${\mathcal E}_{\tiny{\textrm{morph}}}$. Provides each location which obeys the morphology constraint with a measure of the strength of morphology signal. \\
 
\item[$\bullet$] {\it Feature Map} \\ 
For each level of Scale-Space the feature map ${\mathcal F}_{\tiny{\textrm{morph}}}$ is constructed from the Feature Intensity Map $\mathcal{I}_{\tiny{\textrm{morph}}}$ and the Morphology Response Map. This represents local structures as seen on the different scales of the Scale-Space.\\
 
\item[$\bullet$] {\it Scale-Space Map Stack}\\ 
By combining the individual Feature Maps $\mathcal{F}_{L,\tiny{\textrm{morph}}}$ of each level of Scale-Space, the ultimate scale independent map of features is produced, the Scale-Space Map Stack $\Psi$. Each pixel in this map is the maximum value of the corresponding pixels in the Feature maps that make up the Scale-Space stack. \\
 
\item[$\bullet$] {\it Object Maps} \\ 
Astrophysical and Cosmological criteria determine the final Object Maps $\mathcal{O}_{\tiny{\textrm{morph}}}$. These maps are produced by thresholding the Scale-Space Map Stack $\Psi$ according to a criterion translating a feature map of physically recognizable objects. \\
 
\item[$\bullet$] {\it Datapoint identification}\\ 
Datapoints within the feature contours of the object map $\mathcal{O}_{\tiny{\textrm{morph}}}$ are identified.  They are removed from the original dataset at each pass through the feature finding process.  \\
\end{enumerate}

%
\begin{figure*}[t]
  \centering
     \mbox{\hskip -0.5truecm\includegraphics[width=1.0\linewidth]{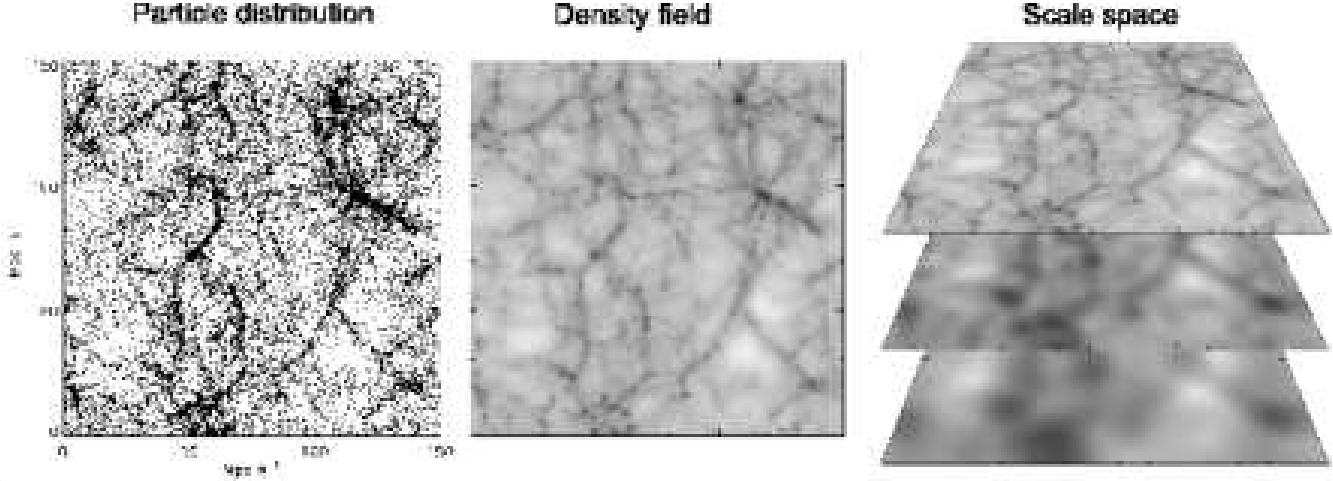}}
     \caption{Scale-space: a particle distribution (left) is translated by DTFE into a density field (centre), 
followed by the determination of the field, by means of filtering, at a range of scales (righthand).}  
\label{fig:scalespace}
\end{figure*}

%
%
\section{Scale Space Technology}
\label{sec:ScaleSpace}
\subsection{Scale-Space Filtering}
The so-called Scale-Space approach to morphology consists simply of calculating and comparing morphology indicators on a variety of scales.  Fundamental in this is the ability to view a given dataset on different scales.  This task is accomplished simply by convolving the original data $f({\vec x})$ with smoothing filters $W$ to produce a smoothed field $f_{\rm S}({\vec x})$:
\begin{equation}
f_{\rm S}({\vec x}) \,=\,\int\,{\rm d}{\vec y}\,f({\vec y})\,W({\vec y},{\vec x})	\nonumber
\end{equation}
The smoothing filter could be any of a number of suitable filters: it is usual, though neither necessary nor optimal, to choose filters based on Gaussian functions.  There are alternatives to this scaling strategy: any form or pyramidal or wavelet transform will have a similar effect.
 
In this paper we generate scaled representations of the data by repeatedly smoothing the DTFE reconstructed density field $f_{\tiny{\textrm{DTFE}}}$ with a hierarchy of spherically symmetric Gaussian filters $W_{\rm G}$ having different widths $R$:
\begin{equation}
f_{\rm S}({\vec x}) =\, \int\,{\rm d}{\vec y}\,f_{\tiny{\textrm{DTFE}}}({\vec y})\,W_{\rm G}({\vec y},{\vec x})\nonumber
\end{equation}
where $W_{\rm G}$ denotes a Gaussian filter of width $R$: 
\begin{equation}
W_{\rm G}({\vec y},{\vec x})\, = \,{1 \over ({2 \pi} R^2)^{3/2}}\, \exp \left(- {|{\vec y}-{\vec x}|^2 \over 2 R^2}\right)\,.
\label{eq:filter}
\end{equation}
A pass of the smoothing filter attenuates structure on scales smaller than the filter width.  
 
The scale-space MMF analysis described in this study involves a discrete number of $N+1$ levels, ${n=0,\ldots,N}$. Following \citep{sato1998} we use a nested hierarchy of filters having widths differing by a factor of $\sqrt{2}$:
\begin{equation}
R_n=(\sqrt{2})^n\,R_0
\end{equation}
The base-scale $R_0$ is taken to be equal to the pixel scale of the raw DTFE density map.  \citet{sato1998} showed that using a ratio of $\approx \sqrt{2}$ between discrete levels involves a deviation of a mere $4\%$ with respect to the ideal case of a continuum of scale-space levels.  \footnote {It is interesting to note also that \cite{marr1980} had already commented on the importance of the $\sqrt{2}$ factor on psycho-visual grounds.} As a retrospective on this research we would argue that, in the context of cosmic structure, the factor of $\sqrt{2}$ is somewhat too coarse.
 
The largest structure that survives this process is determined by the effective width of the filter used in the final smoothing stage. For our purposes it is sufficient to use $n=5$. 
 
We shall denote the $n^{th}$ level smoothed version of the DTFE reconstructed field $f_{\tiny{\textrm{DTFE}}}$ by the symbol $f_n$.
 
The Scale Space itself is constructed by stacking these variously smoothed data sets, yielding the family $\Phi$ of smoothed density maps $f_n$:
\begin{equation}
\label{eq:scalespace}
\Phi\,=\,\bigcup_{levels \; n} f_n 
\end{equation}
A data point can be viewed at any of the scales where scaled data has been generated.  The crux of the concept is that the neighbourhood of a given point will look different at each scale.  There are potentially many ways of making a comparison of the scale dependence of local environment.  We chose here to use the Hessian Matrix of the local density distribution in each of the smoothed replicas of the original data.
 
%
\begin{figure*} 
  \centering
     \mbox{\hskip -0.25truecm\includegraphics[width=1.0\linewidth]{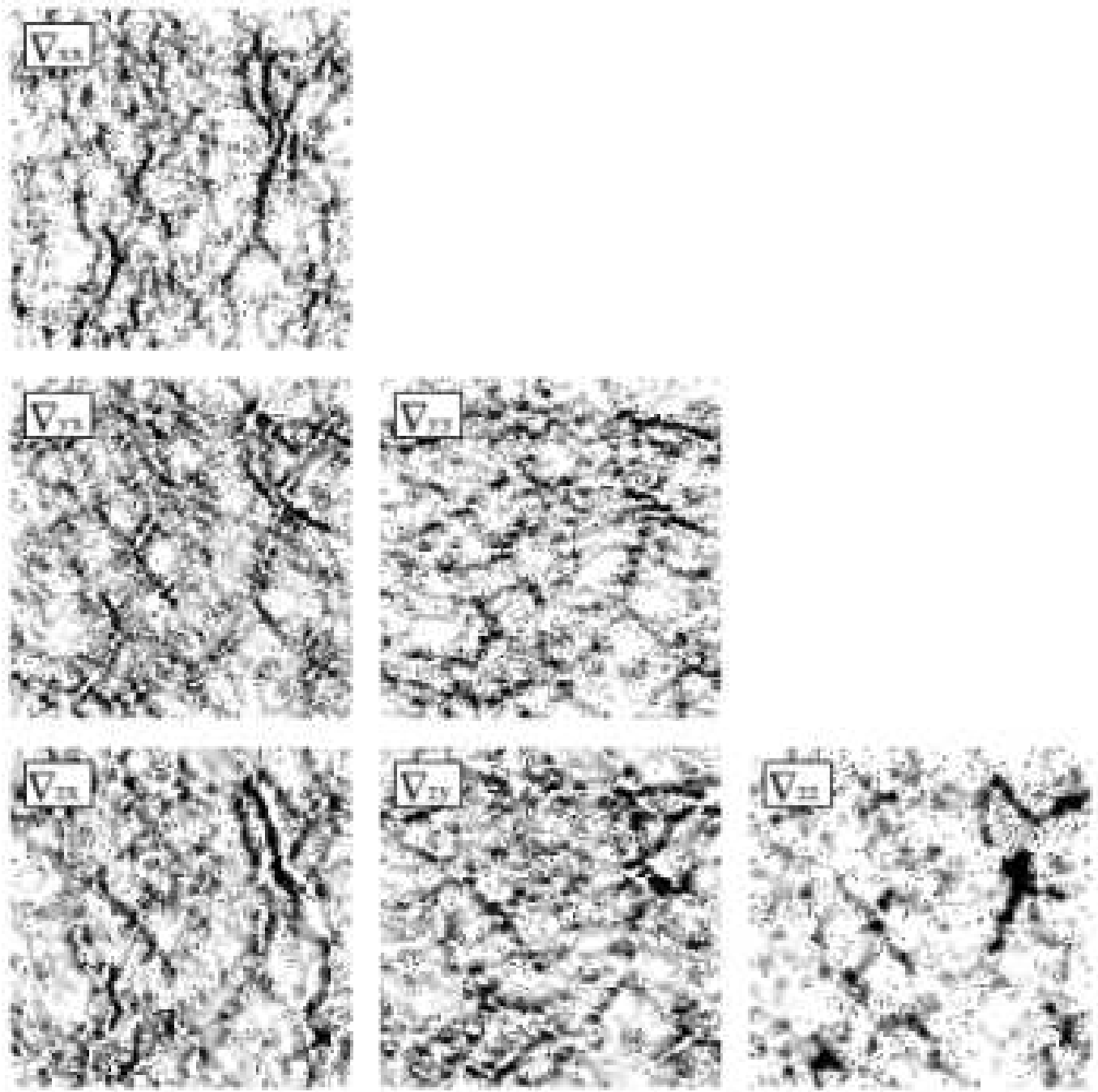}}
     \caption{Maps of the 6 independent components of the (filtered) density field Hessian, ordered by their identity. 
Top row: $\nabla_{11} f$; Central row: $\nabla_{21} f$, $\nabla_{22} f$; Bottom row: $\nabla_{31} f$, $\nabla_{32} f$, $\nabla_{33} f$}  
\label{fig:hessian}
\end{figure*}
%
\begin{figure*} 
  \centering
     \mbox{\hskip -0.25truecm\includegraphics[width=1.0\linewidth]{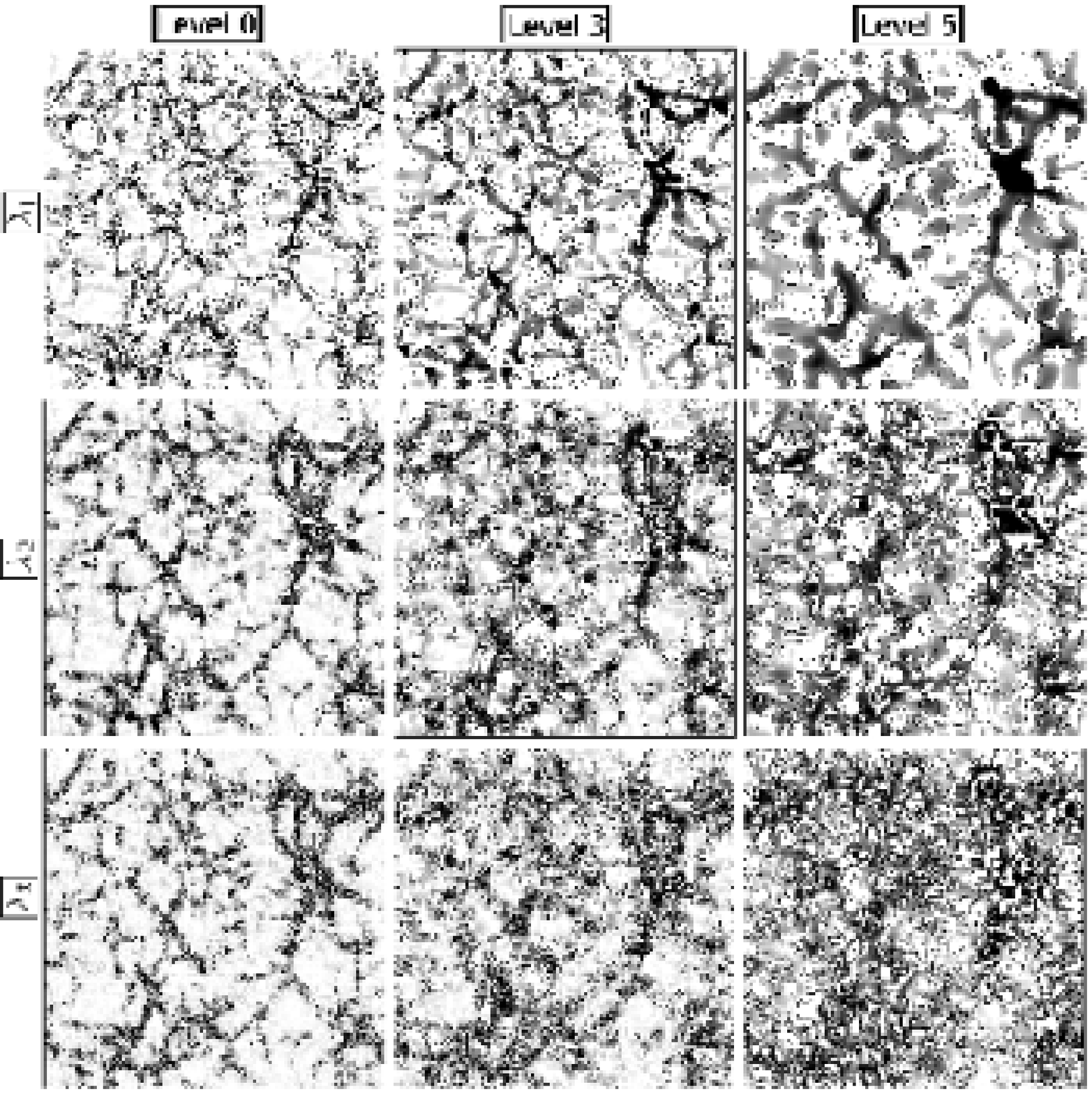}}
     \caption{Maps of the eigenvalues of the Hessian matrix at 3 different scales (levels). From top to bottom: the 3 eigenvalues 
$\lambda_1$, $\lambda_2$ and $\lambda_3$ ($\lambda_1 > \lambda_2 > \lambda_3$). From  left to right: 3 different scales $R_1$
$R_3$ and $R_5$, ($R_1>R_2>R_5$).     Positive values are represented as gray shades in logarithmic scale while negative values are indicated by contour lines also in logarithmic scale.}  
\label{fig:eigenvalue}
\end{figure*}
 
\subsection{The Hessian}
At each point of each dataset in the Scale Space view of the data we can quantify the local ``shape" of the density field in the neighbourhood of that point by calculating, at each point the eigenvalues of the Hessian Matrix of the data values.  
 
We can express the local variations around a point $\vec{x}_0$ of the density field $f(\vec{x})$ as a Taylor expansion:

\begin{equation}\label{eq:taylor_exp_1}
   f(\vec{x}_0 + \vec{s}) = 
              f(\vec{x}_0) + \vec{s}^T \nabla f(\vec{x}_0) +
                \frac{1}{2}\vec{s}^T \mathcal{H} (\vec{x}_0) \vec{s} + ...
\end{equation} 
where
\begin{equation}\label{eq:hessian_1} 
  \mathcal{H} = \left ( \begin{array}{ccc}
                f_{xx} & f_{yx} & f_{zx} \\
                f_{xy} & f_{yy} & f_{zy} \\
                f_{xz} & f_{yz} & f_{zz}
		\end{array} \right )
\end{equation} 
is the Hessian matrix.  Subscripts here denote partial derivatives of $f$ with respect to the named variable.  There are many possible algorithms for evaluating these derivatives.
 
In our case we compute the scale-space Hessian matrices for each level $n$ directly from the DTFE density field, via the convolution 
 
\begin{eqnarray}
&&\frac{\partial^2}{\partial x_i \partial x_j} f_S({\vec x})\,=\,f_{\tiny{\textrm{DTFE}}}\,\otimes\,\frac{\partial^2}{\partial x_i \partial x_j} W_{\rm G}(R_{\rm S})\nonumber \\
&&= \int\,{\rm d}{\vec y}\,f({\vec y})\,\,\frac{(x_i-y_i)(x_j-y_j)-\delta_{ij}R_{\rm S}^2}{R_{\rm S}^4}\,W_{\rm G}({\vec y},{\vec x})
\end{eqnarray} 
where ${x_1,x_2,x_3}={x,y,z}$ and $\delta_{ij}$ is the Kronecker delta. In other words, the scale space representation of the Hessian matrix for each level $n$ is evaluated by means of a convolution with the second derivatives of the Gaussian filter, also known as the Marr (or, less appropriately, ``Mexican Hat'') Wavelet. 
 
In order to properly compare the values of the Hessian arising from the differenlty scaled variants of the data that make up the Scale Space we must use a renormalised Hessian:
\begin{equation}
	\tilde {\mathcal{H}}\,=\,R_{\rm S}^2 \,\mathcal{H}
\end{equation}
where $R_{\rm S}$ is the filter width that has been used ($\sqrt{2}^n R_0$ for level $n$ in our case).  Instead of using this `natural' renormalization, it would be possible to use a scaling factor $R^{2\gamma}$.  Using values $\gamma > 1$ will give a bias towards finding larger structures, while values $\gamma < 1$ will give a bias towards finding smaller structures.
 
%
\begin{figure*}[t] 
  \centering
     \mbox{\hskip -0.25truecm\includegraphics[width=1.0\linewidth]{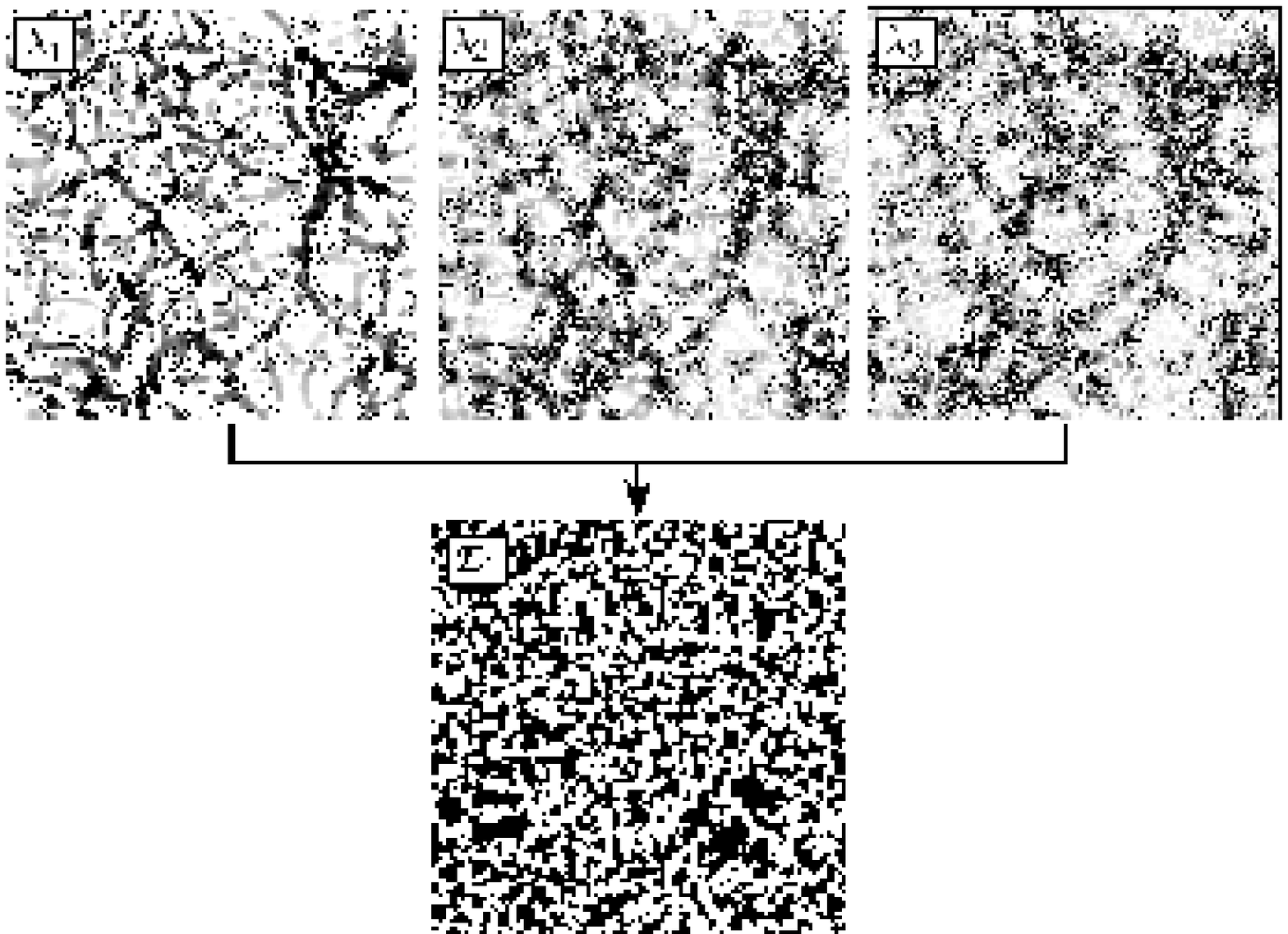}}
     \caption{Morphology Mask ${\mathcal E}$: on the basis of the 3 eigenvalues $\lambda_1$, $\lambda_2$ and $\lambda_3$ at each location we 
determine whether the morphological criterion -- here whether it corresponds to a filament (table 1) -- is valid. If so 
${\mathcal E}=1$, otherwise it is ${\mathcal E}=0$. Top row: maps of the three eigenvalues; bottom row: the Morphology Mask ${\mathcal E}$.}  
\label{fig:lambdaE}
\end{figure*}
 
\subsection{Eigenvalue and Eigenvectors}
The eigenvalues of the Hessian matrix evaluated at a point quantify the rate of change of the field gradient in various directions about each point. The eigenvalues are coordinate independent measures by the components of the second derivatives of the field at each point $\vec{x}_0$.  A small eigenvalue indicates a low rate of change of the field values in the corresponding eigen-direction, and vice versa.
 
We denote these eigenvalues by $\lambda_{a}(\vec{x})$ and arrange them so that $ \lambda_1 \ge \lambda_2 \ge \lambda_3 $:
\begin{eqnarray}
\qquad \bigg\vert \; \frac{\partial^2 f_n({\vec x})}{\partial x_i \partial x_j}  - \lambda_a({\vec x})\; \delta_{ij} \; \bigg\vert  &=& 0,  \quad a = 1,2,3 \\
\mathrm{with} \quad \lambda_1 &>& \lambda_2  >  \lambda_3 \nonumber
\end{eqnarray}
The $\lambda_{i}(\vec{x})$ are coordinate independent descriptors of the behaviour of the density field in the locality of the point $\vec{x}$ and can be combined to create a variety of morphological indicators.  The corresponding eigenvectors show the local orientation of the morphology characteristics.     Note, however, that in this study we do not make use of the eigenvectors.

%
%
\section{Scale-Space Feature Detection and Extraction}
The eigenvalues of the Hessian therefore encode the local morphology of the density field in terms of the curvature components of the local density field in the direction of the corresponding eigenvectors.  Evaluating the eigenvalues and eigenvectors for the renormalised Hessian $\tilde {\mathcal{H}}$ of each dataset in a Scale Space shows how the local morphology changes with scale.
 
With the local curvature and shape encapsulated in the three eigenvalues $\lambda_1$, $\lambda_2$ and $\lambda_3$ of the Hessian, the MMF seeks to identify the regions in space which correspond to a certain morphology and at the scale at which the corresponding morphology signal attains its optimal value. First we set out to select these regions by means of a Morphology Mask. Subsequently we develop a filter-based procedure for assigning at each scale a local weight which is used to select the scale at which the morphology reaches its strongest signal.

\begin{table}
\label{tab:morphmask}
\centering
\begin{tabular} {|c|c|l|}
\hline
\hline
\ &&\\
Structure & $\lambda$ ratios & $\quad$ $\lambda$ constraints \\
\  &&\\
\hline
\ &&\\
Blob     &  $\lambda_1 \simeq \lambda_2 \simeq \lambda_3$ & $\lambda_3 <0\,\,;\,\, \lambda_2 <0 \,\,;\,\, \lambda_1 <0 $  \\
\ &&\\
Line     &  $\lambda_1 \simeq \lambda_2 \gg    \lambda_3$ & $\lambda_3 <0 \,\,;\,\, \lambda_2 <0  $  \\
\ &&\\
Sheet    &  $\lambda_1 \gg    \lambda_2 \simeq \lambda_3$ & $\lambda_3 <0 $    \\
\ &&\\
\hline
\hline
\end{tabular}
\vskip 0.25truecm
\caption{Eigenvalue relationships defining the characteristic morphologies. The
         $\lambda$-conditions describe objects with intensity higher than their
		 local background as clusters, filaments or walls. For voids we would have to reverse the
	     sign of the eigenvalues.}
\end{table}

\begin{figure*}[t]
  \centering
     \mbox{\hskip -0.25truecm\includegraphics[width=1.0\linewidth]{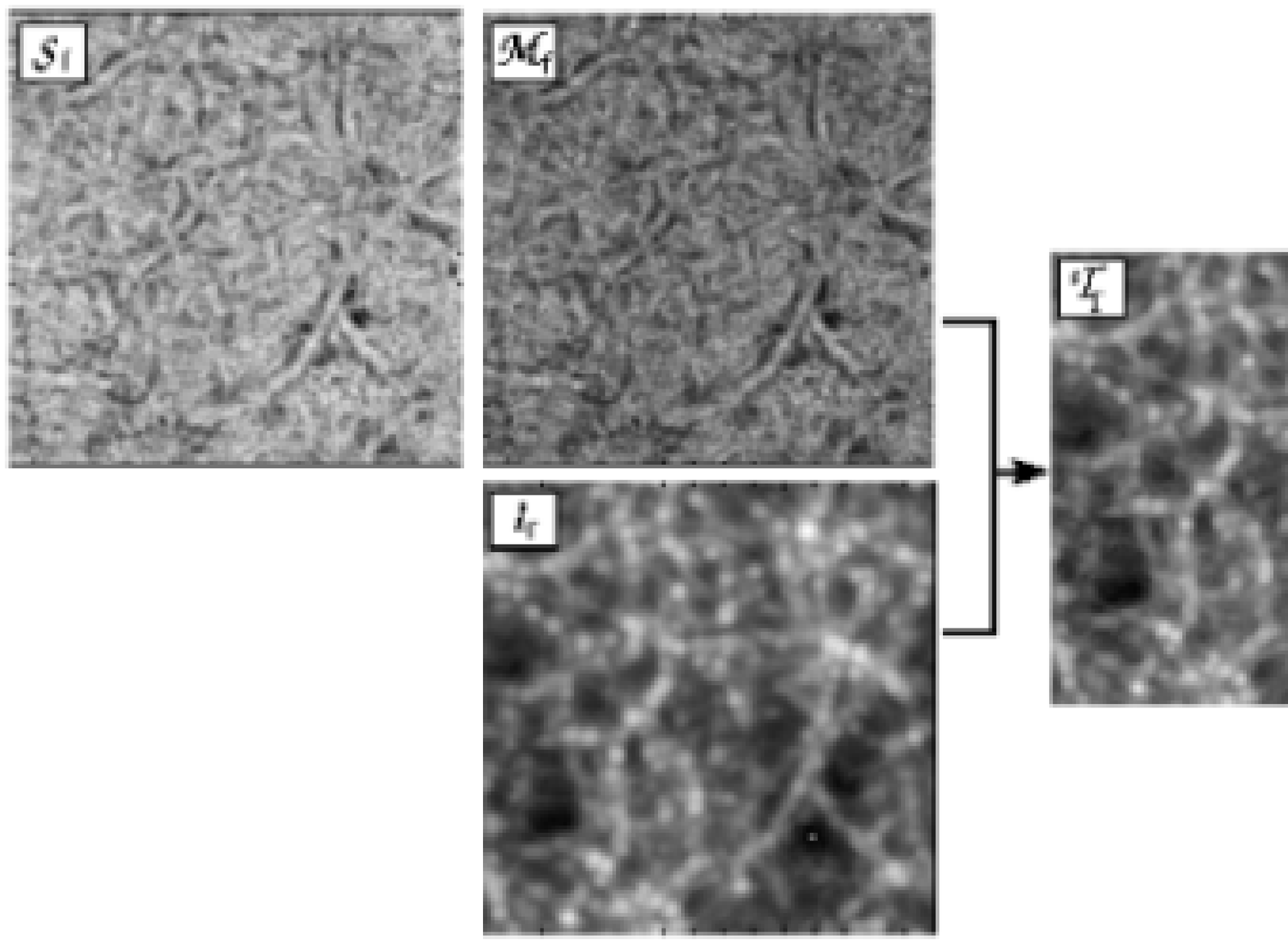}}
     \caption{Morphology Filter ${\mathcal T}$.  The Morphology Response function ${\mathcal M}$ (top centre) is the soft thresholded version of the Shape Significance map ${\mathcal S}$ (left frame), determined from the values of the eigenvalues $\lambda_1$, $\lambda_2$ and $\lambda_3$.  The Morphology Intensity function ${\mathcal I}$ (bottom centre)  is also computed from the $\lambda$'s using equation~({\ref{eq:intensity}}).  Finally, the Morphology Filter ${\mathcal T}$ (right frame) is obtained by combining ${\mathcal M}$ with ${\mathcal I}$.}  
\label{fig:smit}
\end{figure*}
 
\subsection{Morphology Mask: $\mathcal{E}$}
Locally ``spherical'' topology is indicated by all three eigenvalues  being similar in size, and locally ``filamentary'' topology is indicated by having two similar eigenvalues and a negligible third; the direction of the filamentary structure is then in the direction of the eingenvector corresponding to the smallest (insignificant) eigenvalue.  A locally ``sheet-like' structure is characterised by one dominant eigenvalue, its corresponding eigenvector indicating the normal to the sheet. The formal morphology conditions are listed in table~\ref{tab:morphmask}. 
 
There are many ways of using the eigenvalues of the Hessians in the Scale Space representation of the data to identify and demarcate specific types of structure.  Here we start by defining a morphology mask. The {\it Morphology Mask} ${\mathcal E}_{\tiny{\textrm{morph}}}$ is a hard filter which identifies all pixels obeying the morphology and shape condition:
\begin{eqnarray}
\label{eq:morphmask}
     \mathcal{E}_{\tiny{\textrm{morph}}}&\,=\,&
\begin{cases}
\ 1 \qquad \hbox{\rm morphology constraint valid}\nonumber\\
\ \\
\ 0 \qquad \hbox{\rm morphology constraint invalid}\nonumber
\end{cases}
\end{eqnarray}
See figure~\ref{fig:lambdaE} to see how this works.
 

\subsection{Feature shape fidelity: $\mathcal{S}$}
The degree of ``blobbiness'', ``filamentariness'' or ``wallness'' is reflected in the degree to which the inequalities of table 
\ref{tab:morphmask} defining those structures are satisfied.  We would be impressed by a blob in which all three eignevalues were equal - it would look like a spherical lump.  We would be less impressed if there was a factor 3 between the eigenvalues since the blob would then look more like a flattened sausage while not manifestly being a filament or a wall.
 
The following shape indices reflect the strength $\mathcal{S}$ of the classification in terms of the local geometry as characterised by the $\lambda$'s.
\begin{equation} 
\label{eq:significance}
\mathcal{S}_{\tiny{\textrm{morph}}}  = \left\{\begin{array}{ll} 
           \frac{\displaystyle \vert \lambda_3 \vert}{\displaystyle \vert \lambda_1 \vert}  & \quad\textrm{Blob} \\
           \ \\
           \left(1-\frac{\displaystyle \vert \lambda_3 \vert}{\displaystyle \vert \lambda_1\vert}\right) \cdot \frac{\displaystyle \vert \lambda_3 \vert}{\displaystyle \vert \lambda_2 \vert} & \quad\textrm{Filament} \\
           \ \\
           \left(1-\frac{\displaystyle \vert \lambda_3 \vert}{\displaystyle \vert \lambda_1\vert}\right) \cdot \left( 1 - \frac{\displaystyle \vert \lambda_3 \vert}{\displaystyle \vert \lambda_2 \vert}\right) & \quad\textrm{Wall} \\
                      \end{array} \right. 
\end{equation}
It is important to emphasise when using this equation that the values of $\mathcal{S}$ are only meaningful if the relevant inequalities in table \ref{tab:morphmask} are already satisfied.  
 
As a cautionary warning it must be stressed that we cannot identify a point as being part of a locally filamentary structure and assess the significance by using an evaluation of $\mathcal{S}$ that applies to blobs or walls.  Likewise the value of $\mathcal{S}$ cannot be used to assess the relative significance of different types of structure.  This means that the identification of structural elements using this eigenvalue classification scheme must be done cyclically: first find blobs (three roughly equal eigenvalues), then lines (two roughly equal and dominant eigenvalues) and finally walls (one dominant eigenvalue). There are other schemes that are one-pass classifiers.
 
We shall use the symbols $\mathcal{S}_{\tiny{\textrm{blob}}}$, $\mathcal{S}_{\tiny{\textrm{filament}}}$, $\mathcal{S}_{\tiny{\textrm{wall}}}$ to denote the values of $\mathcal{S}$ computed for each kind of feature.
 
\begin{figure*}[t] 
  \centering
     \mbox{\hskip -0.25truecm\includegraphics[width=1.0\linewidth]{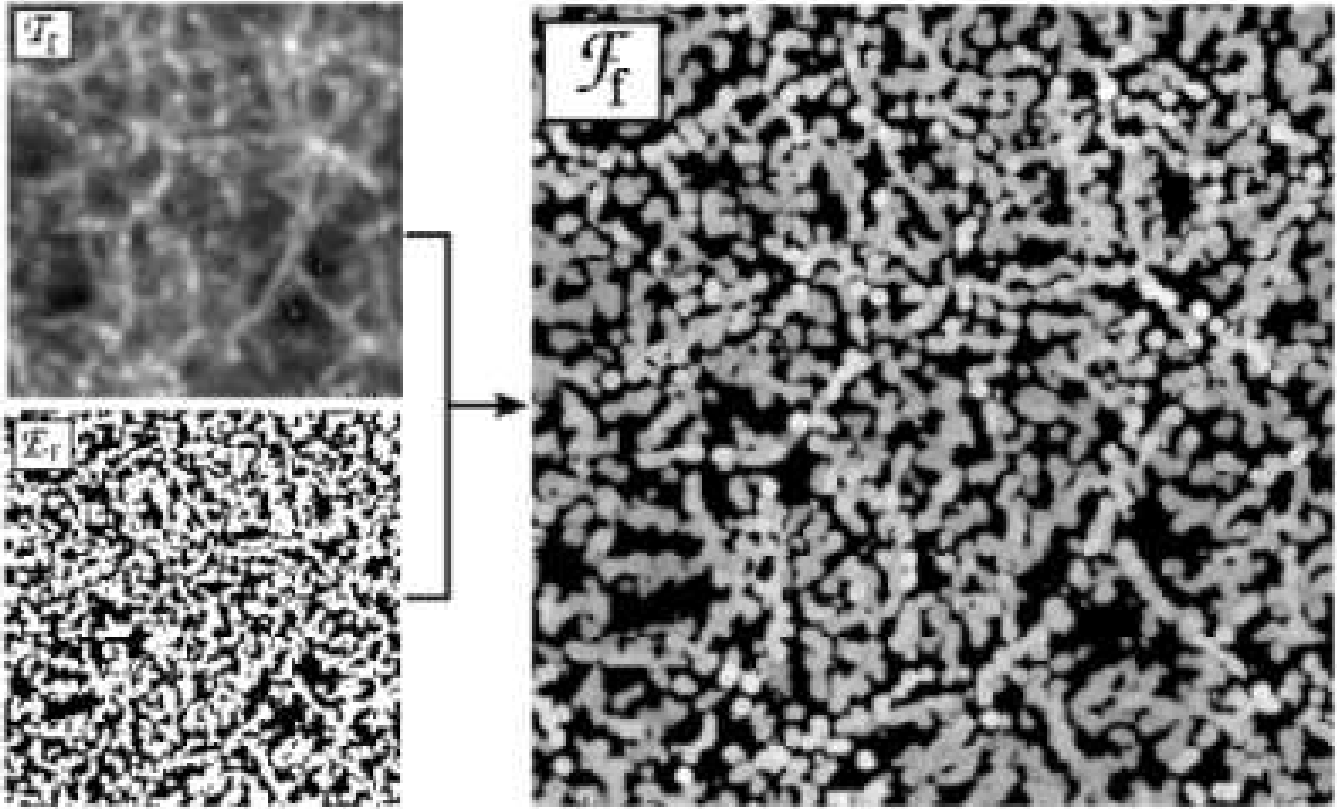}}
     \caption{The Feature Map ${{\mathcal F}}$ (righthand frame) is computed for each scale and is equal to the Morphology Filter ${\mathcal T}$ 
at the locations where the Morphology Mask ${\mathcal E}$ is unity (and nonzero).}  
\label{fig:tef}
\end{figure*}
\subsection{Morphology Response Filter: $\mathcal{M}$}
We shall need a filter that preferentially selects out points where the value of the feature shape parameter $\mathcal{S}$ lies above some threshold.  With this we can tune the aggressiveness of feature-selection. This can be done by defining a morphology measure $\mathcal{M}$ by
\begin{equation}
\label{eq:G_geom_1}
     \mathcal{M}_{\tiny{\textrm{morph}}} = 1 - \exp{ \left( -\frac{\mathcal{S}_{\tiny{\textrm{morph}}}}{2 \beta^2} \right) } 
\end{equation}
where $morph$ = ($blob$, $filament$, or $wall$).   The adjustable parameter $\beta$ tunes the discrimination level of the morphology response filter.  A typical value is $\beta = 0.5$.  Lower values will increase the feature selectivity. Higher values will decrease the selectivity giving feature images with smooth features but contamination from other morphologies.
 
We shall use the symbols $\mathcal{M}_{\tiny{\textrm{blob}}}$, $\mathcal{M}_{\tiny{\textrm{filament}}}$, $\mathcal{M}_{\tiny{\textrm{wall}}}$ to denote the values of $\mathcal{M}$ computed for each kind of feature.
 
Methods of thresholding image data such as equation~(\ref{eq:G_geom_1}) are generally referred to as ``soft thresholding'', as opposed to ``hard thresholding'' in which all values below a critical value are zeroed.  Soft thresholding results in visually more appealing density distributions.  See figure~(\ref{fig:smit}).
\begin{figure*}[t] 
  \centering
     \mbox{\hskip -0.25truecm\includegraphics[width=1.0\linewidth]{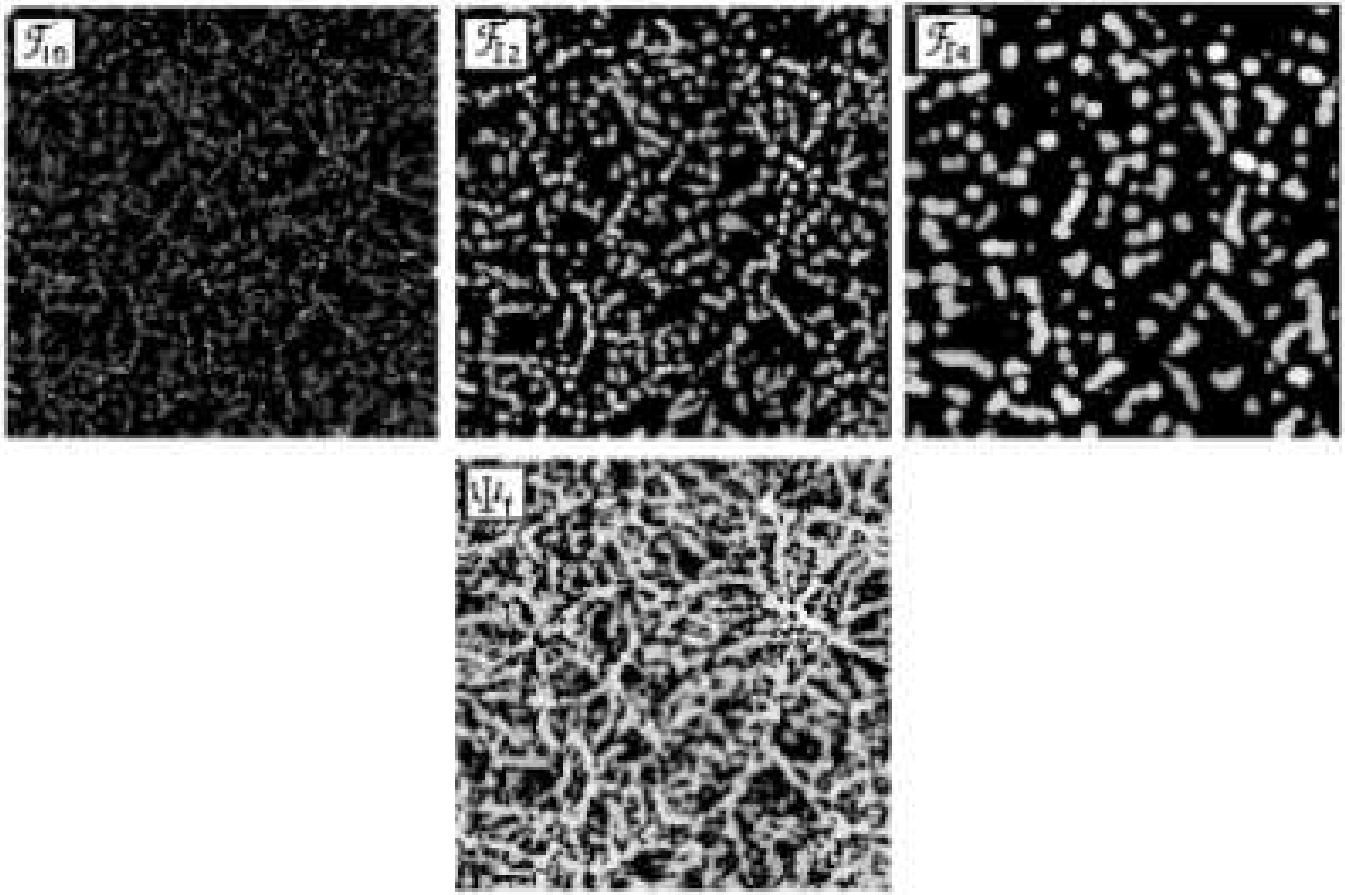}}
     \caption{The Scale Space Map Stack $\Psi$: the formalism selects at any location the scale with the optimal signal 
of the feature map. Depicted are the Feature maps ${\mathcal F}$ for three different scales (top row), and the 
resulting Map Stack $\Psi$ (bottom row), determined over the full range of scales.}  
\label{fig:psi}
\end{figure*}
 
\subsection{Morphology Intensity Map $\mathcal{I}$}
Morphology Intensity is a property of structures that represents how strong the feature is: a filament that is nice and narrow is in some sense more filament-like than one which is rather wide and diffuse.  
The discriminating factor in this case is the magnitude of the eigenvalue $\lambda_2$.  
Note that it would be inappropriate to normalise or non-dimensionalize this relative to some local values such as the sum of the local $\lambda$'s: 
it is the fact of comparing the $\lambda$ values at different spatial locations that discriminates features.  
If, in our example, the value of $\lambda_2$ were roughly constant over the data set, we would not be impressed by any filamentariness.
 
\citet{qian2003} noted that the smallest eigenvalue ($\lambda_3$) will be large only for blobs,  while $\lambda_2$ will be large for blobs and filaments, and $\lambda_1$  for blobs, filaments, and walls. Combining these relations with the $\lambda$ constraints in table \ref{tab:morphmask} we can use the following intensity function:
\begin{equation}
\label{eq:intensity}
\mathcal{I}_{\tiny{\textrm{morph}}}=\left\{\begin{array}{ll} 
\lambda_3  & \textrm{\hspace{0.6cm} Blob} \\
\ \\
\lambda_2  & \textrm{\hspace{0.6cm} Filament} \\
\ \\
\lambda_1  & \textrm{\hspace{0.6cm} Wall} \\
\end{array} \right. 
\end{equation}
The use of this morphology intensity function solves the problem of detecting low-intensity/noisy structures but it introduces another problem: the range of values of $\mathrm{I}_{\tiny{\textrm{morph}}}$ is not well defined within a given interval since it depends on the nature of the density field itself. We therefore normalise its values in the interval $[0,1]$ in order to apply it in a consistent way.
 
There are other posible measures of feature intensity.  \citep{frangi1998} introduced the Frobenius matrix $\sqrt{\lambda_1^2+\lambda_2^2+\lambda_3^2}$ as a measure of second-order structure. However, this measure is biased towards blob-like structures and can produce erroneous signals in the detection of filaments and walls.
 
\subsection{Morphology Filter $\mathcal{T}$}
For each level of the scale space, we can generate a Morphology Filter, $\mathcal{T}$, from the Morphology Intensity Map $\mathcal{I}$ and Morphology Response Filter $\mathcal{M}$.  Formally we can write this as
\begin{equation}
\label{eq:morphfilter}
\mathcal{T}\,=\,\mathcal{I} \otimes  \mathcal{M}
\end{equation}
where the combination operator $\otimes$ simply means that every pixel of the Morphology Intensity Map, $\mathcal{I}$, is multiplied by the value of the corresponding pixel in the Morphology Response Filter $\mathcal{M}$. As described above, these hold information on different aspects of the structural morphology, and by combining them we can hope to improve on the results that would be obtained by using either of them alone. Thus the Morphology Filter has its most significant values at those places where the morphology is close to what we are looking for.

\subsection{Feature Map $\mathcal{F}$} 
This is where, for each level of scale space, we combine information contained in the morphology mask $\mathcal{E}$ and filter $\mathcal{T}$: we select out those regions of ${\mathcal T}$ where the morphology constraint is valid.
 
For each level of the scale space, we can generate a Feature Map, $\mathcal{F}$. The feature map comprises the information 
contained in the Morphology Filter ${\mathcal T}$ and allocates it to the locations contained in the Morphology Mask ${\mathcal E}$.  Formally we can write this as
\begin{equation}
\label{eq:featuremap}
\mathcal{F}\,=\,\mathcal{E} \otimes  \mathcal{T}
\end{equation}
where the combination operator $\otimes$ simply means that every pixel of the Morphology Filter, $\mathcal{T}$, 
is modulated by the mask value ${\mathcal E}$, $1$ or $0$ dependent on whether the morphology constraint is valid at the corresponding location. See figure~(\ref{fig:tef}).
 
\subsection{Scale Space Map Stack $\Psi$}
Each level of the scale space has its Feature Map constructed according to equation (\ref{eq:featuremap}). They must now be combined in order to produce the definitive scale independent map of features, $\Psi$.  We can refer to $\Psi$ as the ``feature stack'' and formally write it as
\begin{equation}
\label{eq:featurestack}
\Psi\,=\,\biguplus_{levels \; n} \mathcal{F}_{n} 
\end{equation}
where the combination operator $\biguplus$ represents a new Feature Map built by combining the individual Feature Maps, $\mathcal{F}_{n}$, of the scale space.  Each pixel of $\Psi$ takes on the maximum value of the corresponding pixel values in the stack of Feature Maps $\mathcal{F}_n$ in the Scale Space. We can write this (for a 3-D map) as
\begin{equation}
\Psi(i,j,k) = \max_{Levels \; n} \;  \mathcal{F}_n(i,j,k)
\label{eq:max_scale_value}
\end{equation}
where $i,j,k$ represent the location of the pixels in the map.  In this way we assign each point of the dataset a value quantifying the degree to which it can be said to be a part of some feature (blob, filament, or wall) on any of the scales investigated by the scale space.  See figure~(\ref{fig:psi}).
 
\begin{figure}[htp]
 \vskip -0.25truecm
 \begin{center}
  \subfigure[Threshold determination for blobs]
	{
	\label{fig:mmf_tau_blobs}
     \mbox{\hskip -0.87truecm\includegraphics[width=0.63\linewidth]{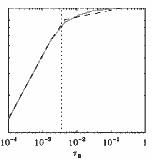}}
	}
  \subfigure[Threshold determination for filaments]
	{
    \label{fig:mmf_tau_filaments} 
     \mbox{\hskip -0.0truecm\includegraphics[width=0.63\linewidth]{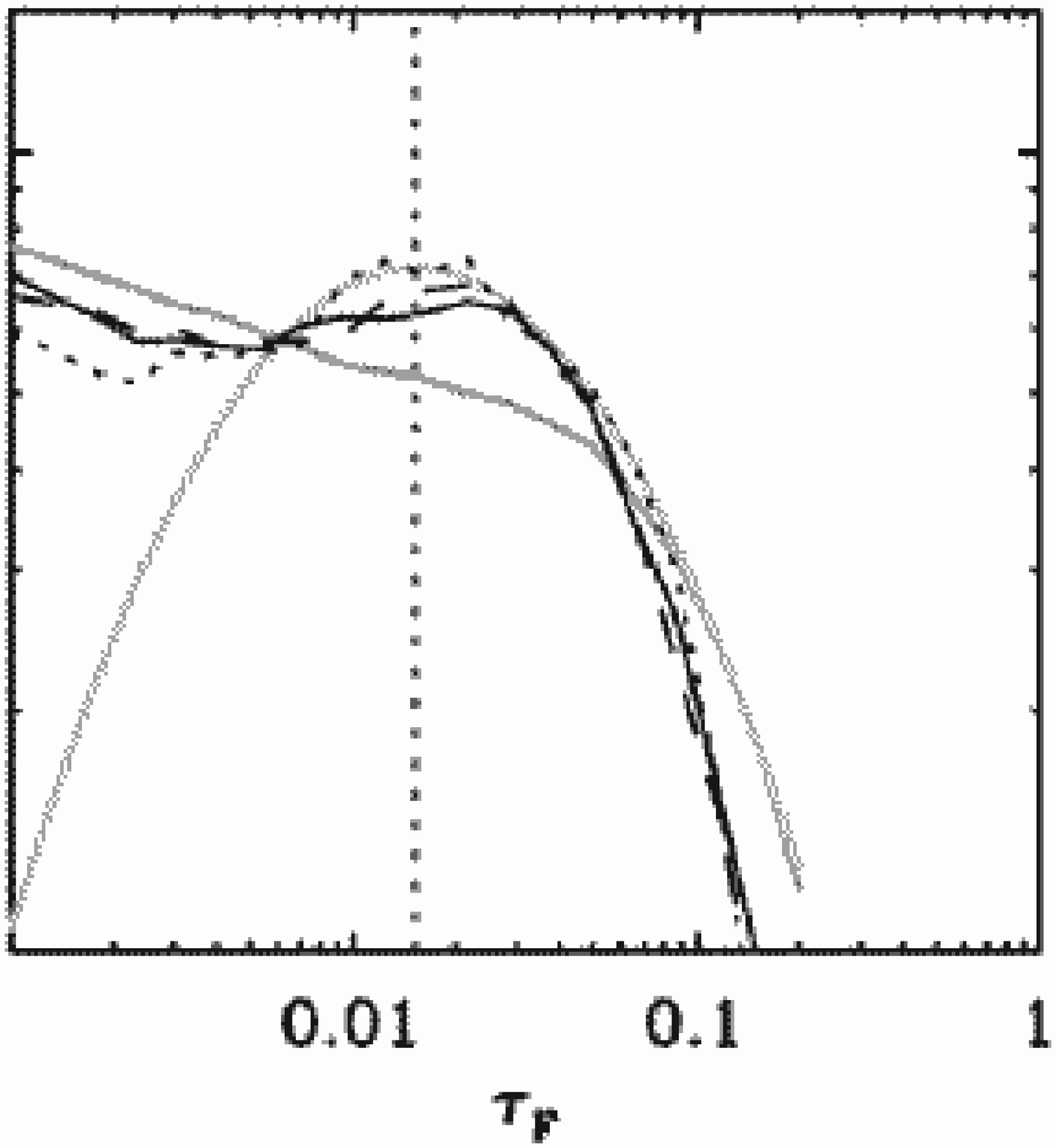}}
	}
  \subfigure[Threshold determination for walls]
	{
    \label{fig:mmf_tau_walls}
     \mbox{\hskip -0.0truecm\includegraphics[width=0.63\linewidth]{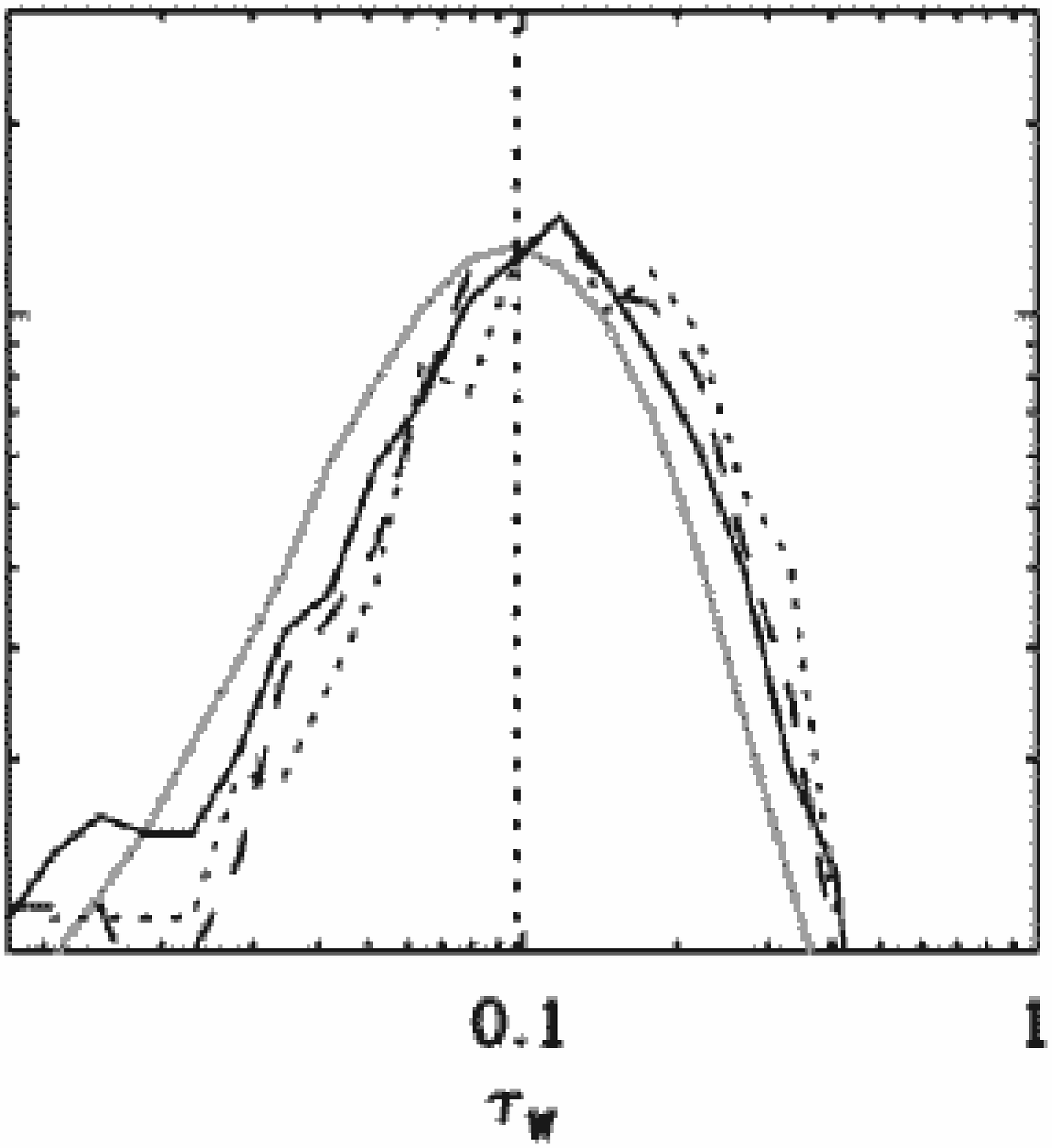}}
	}
 \end{center}
\end{figure}
\begin{figure}[t]
\caption{Thresholds for feature isolation based on the feature erosion criterion.  The selected value is shown as a dotted vertical line.  The object count to the right of the line declines due to erosion.}
\label{fig:Thresholds}
\end{figure}

\subsection{Assigning Points to Features}
The Scale Space Map Stack $\Psi$ has to be thresholded in order to identify the most significant features.  Tiis will be discussed in detail in Section~(\ref{sec:FeatureDetection}).  It is at this point that we see astronomical input by requiring that the sought-after structure correspond to some structure that we would recognise.
 
Given the Scale Space Map Stack $\Psi$ for a given feature (blobs, filaments or walls), we can assign each particle of the original dataset to the specific feature identified in the Scale Space Map Stack.

%
%
\section{Cosmological Feature Detection:\\ \ \ \ \ Threshold definition}
\label{sec:FeatureDetection}

\subsection{Texture Noise}
The final stage of each cycle of the analysis is the thresholding of the scale space map stack in order to identify individual objects that are being sought in that cycle.  Without the thresholding the maps are noisy and over-structured: we can refer to this as as ``\textit{texture noise}''.  This texture noise is simplest removed by applying a simple threshold to the processed maps.  There is a potential problem in applying a simple threshold:  it is necessary to determine a threshold that removes texture noise (however that is determined) while leaving the sought-after features intact.  

\subsection{Object Erosion Threshold}
We set the thresholds for each feature to the value such that raising the threshold higher would start eroding objects and decrease their number.  In other words, the threshold value is set so that the object count is maximised while at the same time texture noise is eliminated.

\subsection{Identifying blobs}
We use $\tau_B$ to denote the value of $\Psi$ above which a pixel is considered as part of a blob.  Figure \ref{fig:mmf_tau_blobs} plots the number of objects detected above each value of the threshold, $\tau_B$.  

For blob finding the thresholding is quite straightforward. At very low threshold, there will be many round objects (the eigenvalue criterion fixes their shape) of which only a small fraction will be the blobs we are seeking.  As the threshold is raised from zero, the noise and the the less significant blobs are eliminated.  There comes a point when the threshold stops annihilating these small, less significant, blobs and simply starts eroding the large blobs.  This is the point where we define out optimal threshold.  The dotted vertical line indicates the best value of $\tau_B$.   

If we plot a graph of the fraction of the sample volume occupied by below-threshold blobs against the threshold we obviously find a monotonic curve that rises from zero to one.  This is shown in Figure \ref{fig:mmf_tau_blobs} where we see a two power-law behaviour with a break marking where the transition from texture noise annihilation to simple blob erosion takes place.

\subsection{Identifying Filaments and Walls}
For filament and wall finding we again choose to threshold the distributions, but this time we decide on the optimal value of the threshold on the basis of the population curve of features defined at each threshold value.  

\subsubsection{Filaments}

We use $\tau_F$ to denote the value of $\Psi$ above which a pixel is considered as part of a filament.  Figure \ref{fig:mmf_tau_filaments} plots the normalised number of objects detected for each value of the threshold, $\tau_F$.  

The explanation for the shape of this curve is as follows.  The low threshold (small-$\tau_F$) objects are largely due to texture noise: the number of these declines as the threshold increases.  When real filamentary features appear the number of detections increases with $\tau_F$ to reach a maximum.  This is because at lower thresholds the features tend to percolate, so that raising the threshold breaks the structure up into a greater number of filamentary objects.  As the threshold rises further the filaments are eroded and get rarer. The point at which filament erosion starts to act is taken as the optimal value of  $\tau_F$. This is indicated by the dotted line in the figure.

\subsubsection{Walls}

We use $\tau_W$ to denote the value of $\Psi$ above which a pixel is considered as part of a wall. Figure \ref{fig:mmf_tau_walls} plots the normalised number of objects detected for each value of the threshold, $\tau_W$.

The threshold for defining walls is determined in the same way as for filaments.  Note, however, that the particles classified as lying in blobs and filaments have been removed in previous cycles of the analysis so there is no longer a significant texture noise component. As the threshold is varied there is a peak in the number of walls that are found.  At thresholds below this critical value the walls join up and percolate, eventually leaving one vast percolating structure.  At higher threshold values walls are eroded and eventually destroyed.  The dotted vertical line indicates the best value of $\tau_W$

\subsection{Pseudo-code}
We have described the process of constructing a Feature Map and identifying features in that map.  However there is a complication that arises in practise because both the Intensity Map and the Morphology Filter are built on a hierarchy of $\lambda$ values.  In the case of the Morphology Filter, the different $\lambda$'s come in through equations (\ref{eq:significance}) and (\ref{eq:G_geom_1}).    In the case of the Intensity Map, different $\lambda$'s  define the strength of different features as described in equation (\ref{eq:intensity}).

The analysis cycle can be expressed in pseudo-code (see accompanying code in next column). In this form of pseudo-code, keywords 
(which correspond to class methods in object oriented programming) are in boldface.

The nature of the hierarchy is such that we have first to identify blobs, remove them from the sample, then identify filaments, and after removing them from the sample finally identify the walls.  This arises because data points in blobs are defined by having three significant eigenvalues, data points in  filaments are defined by having two significant eigenvalues, and data points in walls have only one significant eigenvalue.  Identifying a filament before eliminating blobs would not work since the blobs would be more strongly detected.

%
%

\bigskip
\bigskip
\scriptsize 
\begin{alltt}
\hrule
  \textbf{get} PointSet
  \textbf{set} Feature = Blobs
 
  \textbf{:} Map_Feature
 
  \textbf{resample} PointSet \textbf{to} Mesh \textbf{using} DTFE
  \textbf{construct} ScaleSpace Hierarchy 
 
  \textbf{for each} Level \textbf{in} ScaleSpace
     \{{
     \textbf{build} Hessian Eigenvalue Maps
    
     \textbf{build using} Eigenvalue Criteria \textbf{for} Feature
        \{{
        Morphology Mask, \(\mathcal{E}\)
        Feature Shape Fidelity, \(\mathcal{S}\)
        Morphology Response Filter, \(\mathcal{M}(\mathcal{S})\)
        Feature Intensity Map, \(\mathcal{I}\)
        \}}
 
    \textbf{generate}
        \{{
        Morphology Filter, \(\mathcal{T} = \mathcal{I} \otimes \mathcal{M}\)
        Feature Map, \(\mathcal{F} = \mathcal{E} \otimes \mathcal{T}\)
        \}}
    \}}

  \textbf{stack} ScaleSpace Feature Maps, \(\Psi = \biguplus \mathcal{F} \)
  \textbf{threshold} Feature Maps \textbf{using} Feature Threshold Method 
 
  \textbf{in} thresholded regions
      \{{
      \textbf{identify} Points 
      \textbf{publish} Points
      \textbf{remove} Points \textbf{from} PointSet
      \}}
 
    \textbf{if} Feature = Blobs
        \textbf{set} Feature = Filaments
    \textbf{else if} Feature = Filaments
        \textbf{set} Feature = Walls
    \textbf{else}
         \textbf{quit}

  \textbf{goto} Map_Feature
  
\hrule
\end{alltt}
\normalsize

\bigskip

\section{Areas of further development}
\label{sec:developments}
The methodology we have presented is very simple, yet, as we shall see, it is highly effective in differentiating the three main structural features that make up the cosmic web.  The following section will test the methodology against a sample with controlled clustering: the Voronoi model, and present results for an N-Body simulation.  Before going on to that analysis it is worth making a few remarks about some details of our procedure that might be enhanced.

Our use of isotropic Gaussian filters is perhaps the most important limiting factor in this analysis.  The largest filter radius which is chosen is substantially smaller than the lengths of the typical filaments.  Only the shorter filaments will get isotropised and they are ``lost" since they make no contribution in the scalespace stack.  Our algorithm is indeed a long thin filament finder.  The main side-effect of the Gaussian smoothing is to make the profile (perpendicular to the filament) of the sharper (narrow) filaments Gaussian.  A narrow filament having high density contrast will, under linear Gaussian smoothing, spill over into the large scales at a variety of thresholds and it will appear to be fatter than it really is.  This latter problem is a consequence of applying simple linear filters: it is generally overcome within the scale space context by using nonlinear filters or by using wavelets \citep{martinez2005,saar2007}

Another area for improvement is to use the eigenvectors as well as the eigenvalues themselves.  Here we have simply relied on the relative magnitudes of the eigenvalues as indicators of curvature morphology.  Had the eignevectors themselves been uncorrelated we might have concluded that there was structure when in fact there was only noise: the eigenvector correlations are good indicators of noise.

A third area for improvement would be to use anisotropic smoothing filters.  This leads us into another related approach to this problem: the use of nonlinear diffusion equations to locate structural features.  This will be the subject of another article later on.

%
%
\section{Voronoi Clustering models}
\label{sec:Voronoi}
To test and calibrate the Multiscale Morphology Filter we have
applied the MMF to a set of four {\it Voronoi Element Models}.
These models combine the spatial intricacies of the cosmic web with
the virtues of a model that has a priori known properties. They are particularly
suited for studying systematic properties of spatial galaxy distributions confined
to one or more structural elements of nontrivial geometric spatial patterns.
The Voronoi models offer flexible templates for cellular patterns, and they are easy to
tune towards a particular spatial cellular morphology. In the case of the Voronoi models
we have exact quantitative information on the location, geometry and identity of the
spatial components against which we compare the outcome of the MMF analysis.

\subsection{Voronoi Models}
{\it Voronoi Clustering Models} are a class of heuristic models for
cellular distributions of matter \cite{weygaertphd1991,weygaert2002}. They use
the Voronoi tessellation as the skeleton of the cosmic matter distribution,
identifying the structural frame around which matter will gradually assemble
during the emergence of cosmic structure \cite{voronoi1908,okabe2000}. The interior of
Voronoi {\it cells} correspond to voids and the Voronoi {\it planes} with sheets of
galaxies. The {\it edges} delineating the rim of each wall are
identified with the filaments in the galaxy distribution.  What is
usually denoted as a flattened ``supercluster'' will comprise an
assembly of various connecting walls in the Voronoi foam, as elongated
``superclusters'' of ``filaments'' will usually consist of a few
coupled edges. The most outstanding structural elements are the {\it
vertices}, corresponding to the very dense compact nodes within the
cosmic web, rich clusters of galaxies.

A more detailed description of the model construction may be found in Appendix~\ref{app:vorclustform}.
We distinguish two different yet complementary approaches, {\it Voronoi Element Models} and
{\it kinematic Voronoi models}.  

Simple Voronoi models confine their galaxy distributions to one of the
distinct structural components of a Voronoi tessellation:
\begin{enumerate}
\item[$\bullet$] \emph{Field}\\ Particles located in the {\it interior of Voronoi cells}
(and thus randomly distributed across the entire model box)
\item[$\bullet$] \emph{Wall}\\ Particles within and around the {\it Voronoi walls}.
\item[$\bullet$] \emph{Filament}\\Particles within and around the {\it Voronoi edges}.
\item[$\bullet$] \emph{Blobs}\\ Particles within and around the {\it Voronoi vertices}.
\end{enumerate}
Starting from a random initial distribution of $N$ points, these {\it galaxies} are projected onto
the relevant wall, edge or vertex of the Voronoi cell in whose interior they are initially located.

%
%
\begin{figure*}[htbp]
  \centering
  \includegraphics[width=1.0\linewidth,angle=0.0]{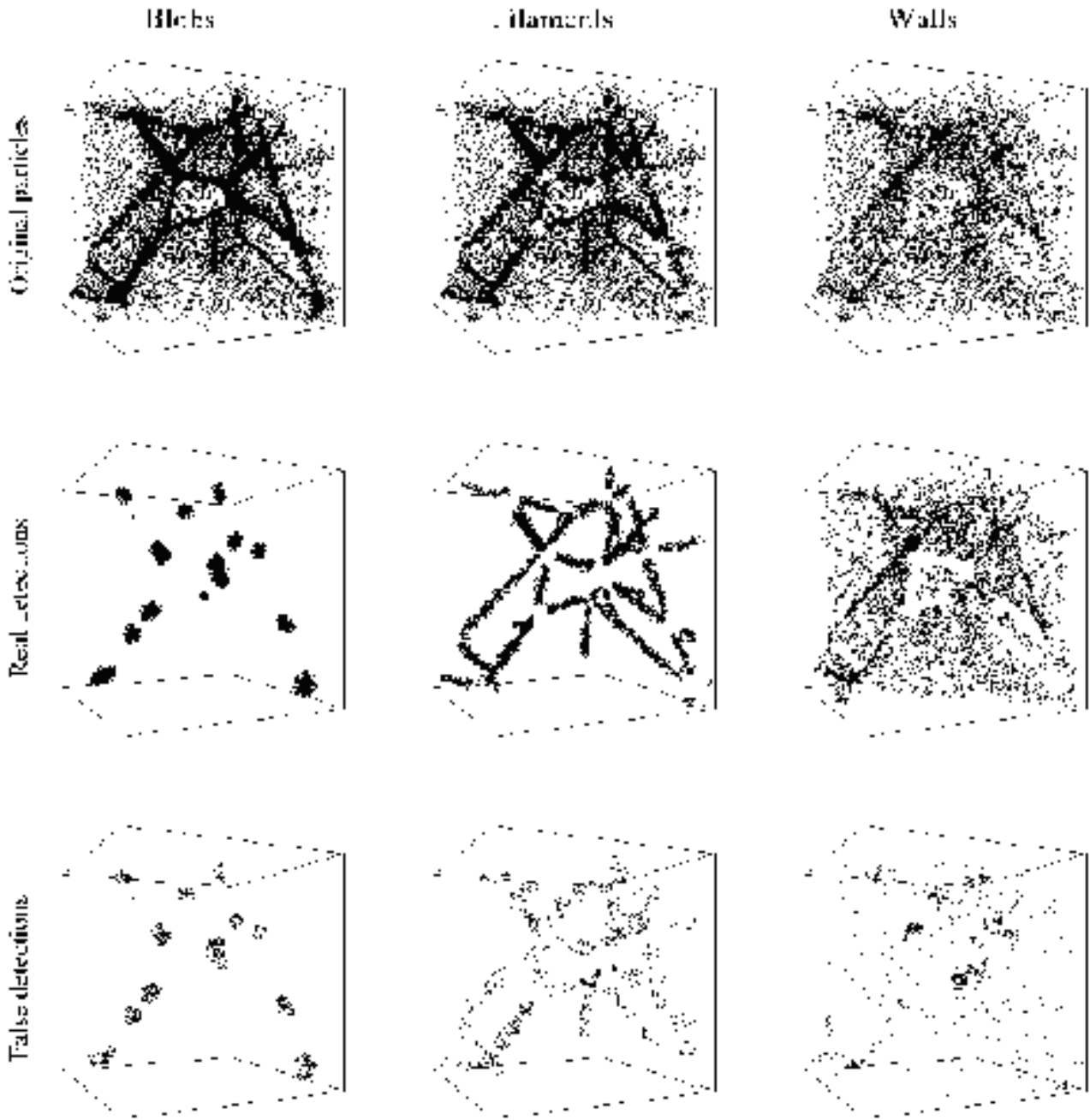}
    \caption{Recovered particles in Blobs, Filaments and Walls from a voronoi particle distribution.
              Particles inside blobs are detected (left), at 90/15 percent real/false detections.
	      From the new blob-free distribution we detect particles in filaments (center)
	      at 90/10 percent real/false detections. Finally the blob-filament-free distribution
	      is used to find the particles inside walls (right) at 80/10 percent real/false detections."
	      }
  \label{fig:particles_recovered_voronoi}
\end{figure*}

For our study we generated four different Voronoi clustering models, labelled as A, B, C and D.
They are all based upon a Voronoi tessellation generated by $M=53$ nuclei distributed within a box of size
$L=100\hmpc$. The models are {\it composite Voronoi Element Models} and consist of the superposition
of galaxies located in field, walls, edges and vertices of a Voronoi tessellation. Our four test models
contain $N=32^3$ galaxies. The fraction of galaxies in the various components is a key parameter of the
model, and is specified in Table~\ref{tab:voronoi_models}. In and around the walls, edges and
vertices the galaxy distribution follows a radial Gaussian density profile, with scale factors
$\sigma_{\rm W}=1.0\hmpc$, $\sigma_{\rm F}=1.0\hmpc$ and $\sigma_{\rm B}=0.5\hmpc$.

\subsection{MMF Processing}

\begin{table}
\label{tab:voronoi_models}
\begin{center}
\begin{tabular} {|l|c|c|c|c|}
\hline
\hline
\ &&&& \\
Model & \% blob & \% filament & \% wall & \% field  \\
\ &&&& \\
\hline
\ &&&& \\
A     & 40   & 30       & 25   & 5 \\
B     & 43   & 17       & 32   & 8 \\
C     & 23   & 37       & 33   & 7 \\
D     & 27   & 23       & 42   & 8 \\
\ &&&& \\
\hline
\hline
\end{tabular}
\medskip
\caption{Voronoi Clustering Models. Percentage of galaxies/points in the various
morphological elements of the model.}
\end{center}
\end{table}

A considerable virtue of the Voronoi clustering models is that it is a priori known which galaxies
reside in the various morphological components of the Voronoi test models. This allows an
evaluation of the absolute performance of the MMF and other morphology detection techniques
by determining the fraction of the galaxies which are correctly identified as vertex, filament
and wall galaxy.

For each Voronoi model we computed the DTFE density field from the particle distribution
and applied the MMF. Following our previously described scheme, we first identified the blobs from the
complete particle distribution. After removal of the blob particles, the filaments
are found. Following the equivalent process for the filaments, the last step of
the MMF procedure concerns the identification of the wall particles. The remaining
particles are tagged as field particles.

Figure~(\ref{fig:particles_recovered_voronoi}) shows the outcome of the MMF applied to
Voronoi Model C. Visually, the resemblance between real and MMF identified blob,
filament and wall particles is remarkably good. The second row of panels shows the real
detections of MMF: MMF clearly manages to identify all clusters, filaments and
even the more tenuous walls in the weblike galaxy distribution. The false detections do
appear to have a somewhat broader spatial distribution than those of the corresponding
real detections. Most of them reside in the boundary regions of the blobs, filaments and
walls: they are mainly an artefact due to the fact that the effective extent of the MMF
morphology masks is slightly larger than the intrinsic extent of the Voronoi components.
Fine-tuning of the filter scales (eq.~\ref{eq:filter}) is a potential solution for
curing this artefact.

\begin{figure*}
  \centering
  \includegraphics[width=\linewidth,angle=0.0]{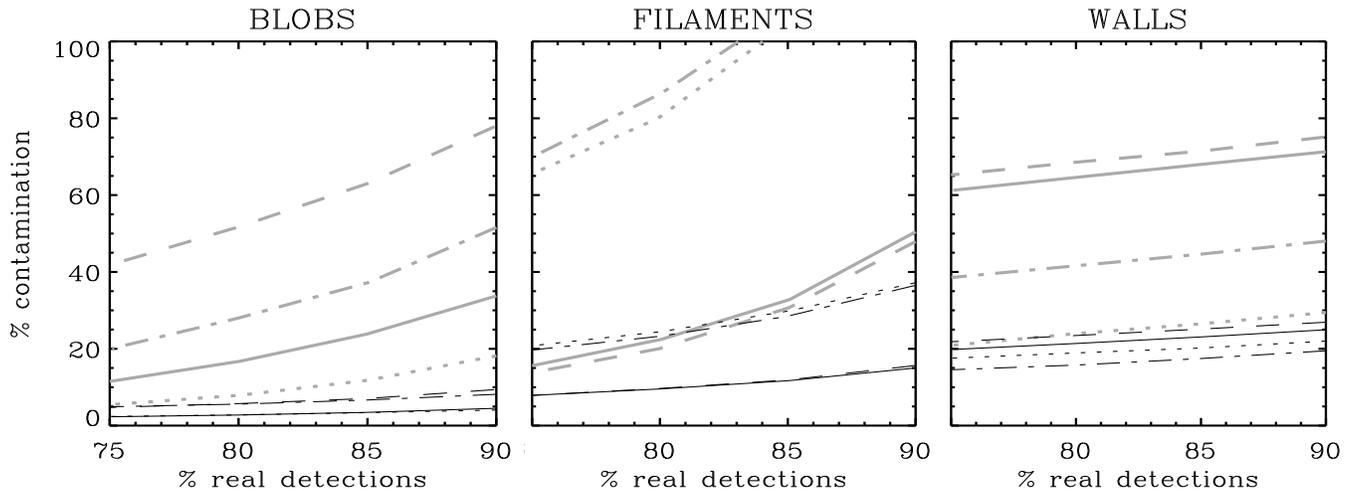}
    \caption{Reals versus false detections for different voronoi models (see table \ref{tab:voronoi_models})
             (A: solid, B:dotted, C:dashed, D:dotted-dashed) for blobs (left), filaments (center) and
             walls (right). We applied the MMF (black) and simple density thresholding (grey) in
             order to compare both methods.
             }
  \label{fig:real_vs_false}
\end{figure*}

\subsection{Detection rate and Contamination}
The detection rate of blob, filament and walls galaxies is determined and
defined as follows. The galaxies in an MMF blob, filament or wall Map Stack $\Psi$
which are genuine Voronoi cluster, filament or wall galaxies are tagged as {\it real}
detections. A galaxy detected by one of the three map stacks $\Psi_b$, $\Psi_f$ or $\Psi_w$
intrinsically belonging to another morphological component is considered
a {\it false detection}. For instance, a filament galaxy detected by $\Psi_b$
is a {\it false} blob galaxy.

The main tunable parameters for optimizing the number of detected galaxies are
blob, filament and wall threshold values, $\tau_b$, $\tau_f$ and $\tau_w$.
By lowering the blob threshold level $\tau_b$, defined through a
regular density threshold (see sect.~\ref{sec:FeatureDetection}), the number of
MMF detected blob galaxies increases. The same holds for adjusting the filament
and wall thresholds, in terms of the lowering of the $\Psi_f$ and $\Psi_w$
levels. The galaxies detected by MMF include both {\it real} and
{\it false} detections. As the threshold levels are adjusted the
number of both will tend to increase.

The {\it detection rate} at a given threshold level is the fraction of genuine
blob, filament or wall galaxies which have been detected by the MMF. Ideally
one would want to trace them all and have a 100\% detection rate, in practice
this is set by the the applied threshold. Based upon the 1-1 relation between
$\tau_b$, $\tau_f$ and $\tau_w$ on the one hand and the corresponding blob, filament
and wall detection rate on the other we use the detection rate as threshold
parameter.

The ratio of the corresponding number of {\it false} blob galaxies to the
{\it total} number of genuine blob galaxies is the blob {\it contamination rate}
rate. The filament and wall {\it contamination rate} are defined in a similar
way. Because a lowering of the threshold levels will result in a larger
number of detections, both {\it real} and {\it false}, the {\it contamination}
rate will be an increasing function of the detection rate. Note that the
{\it contamination rate} may exceed $100\%$ in the case the number of false
detections exceeds that of the total number of genuine (blob, filament or
wall) galaxies.

\subsection{Comparison}
We compare the MMF segmentation of the Voronoi models in blobs, filaments and walls
with that achieved by a more direct criterion, that of a straightforward {\it density
threshold} on the DTFE density field.  We assign the label ``DTC" to this n\"aive procedure. 

Each of the morphological elements are identified with a particular (disjunct) range of density values. Blobs, ie. clusters, are identified with the highest density values. Filaments are associated with more moderately high density values. Walls follow with density values hovering around unity to a few, while the field/voids may include densities down to a zero value. This approach has frequently been used to discriminate between virialized haloes and the surrounding matter distribution, and has even been used in an attempt to define filamentary or planar features \cite{dolag2006}. However, it seriously  oversimplifies and distorts the perceived structure of the cosmic web. (This is presumably because filaments and walls differr in density and have significant internal density structure. The simplistic density threshold approach does not reflect the reality of the structure: the range of densities in filaments overlaps with densities in walls and even with those of the outskirts of clusters. \citep{hahn2007} reach similar conclusions.

\subsection{Test results}
Figure~\ref{fig:real_vs_false} compares the {\it contamination rate} as a function
of the {\it detection rate} for the four different Voronoi models. The A,B, C and
D models are distinguished by means of line style. The black lines relate to the MMF detections,
the grey lines show the results of the equivalent DTC procedure. We find the following:
\begin{enumerate}
\item[$\bullet$] For all models, and for all morphologies, the MMF procedure is clearly
superior to the DTC detections in suffering significantly lower contamination rates.
\item[$\bullet$] The MMF contamination is least for the blob detections. The
filament contamination is lower than the wall contamination for models with many
intrinsic filament galaxies (A and C). For models B and D, containing more
wall galaxies, the situation is the reverse. The same holds true for the DTC
detections, be it much more pronounced and less favorable wrt. the MMF detections.
\item[$\bullet$] The MMF and DTC blob contamination rate is more favorable for the
A and B models. Both models contain a relatively high fraction of blob galaxies.
\item[$\bullet$] The DTC blob contaminations are surprisingly bad, given that
clusters are compact objects of high density with sharply defined boundaries.
\item[$\bullet$] The filament contamination rate is worse for models B and D,
both marked by a relatively low amount of intrinsic filament galaxies. This is true
for both DTC and MMF.
\item[$\bullet$] The DTC contamination is extremely bad for models B and D, quickly
exceeding $100\%$. This reflects the huge overlap in density range of filaments and
other morphologies resulting in a systematic inclusion of particles belonging
to other morphologies.
\item[$\bullet$] For the MMF procedure there is a clear correlation between the
intrinsic wall galaxy fraction and the contamination rate: model D has the highest
number of wall galaxies and the lowest contamination. This is not true for DTC.
\end{enumerate}

%
%
\begin{figure*}[htp]
  \centering
  \includegraphics[width=0.95\linewidth,angle=0.0]{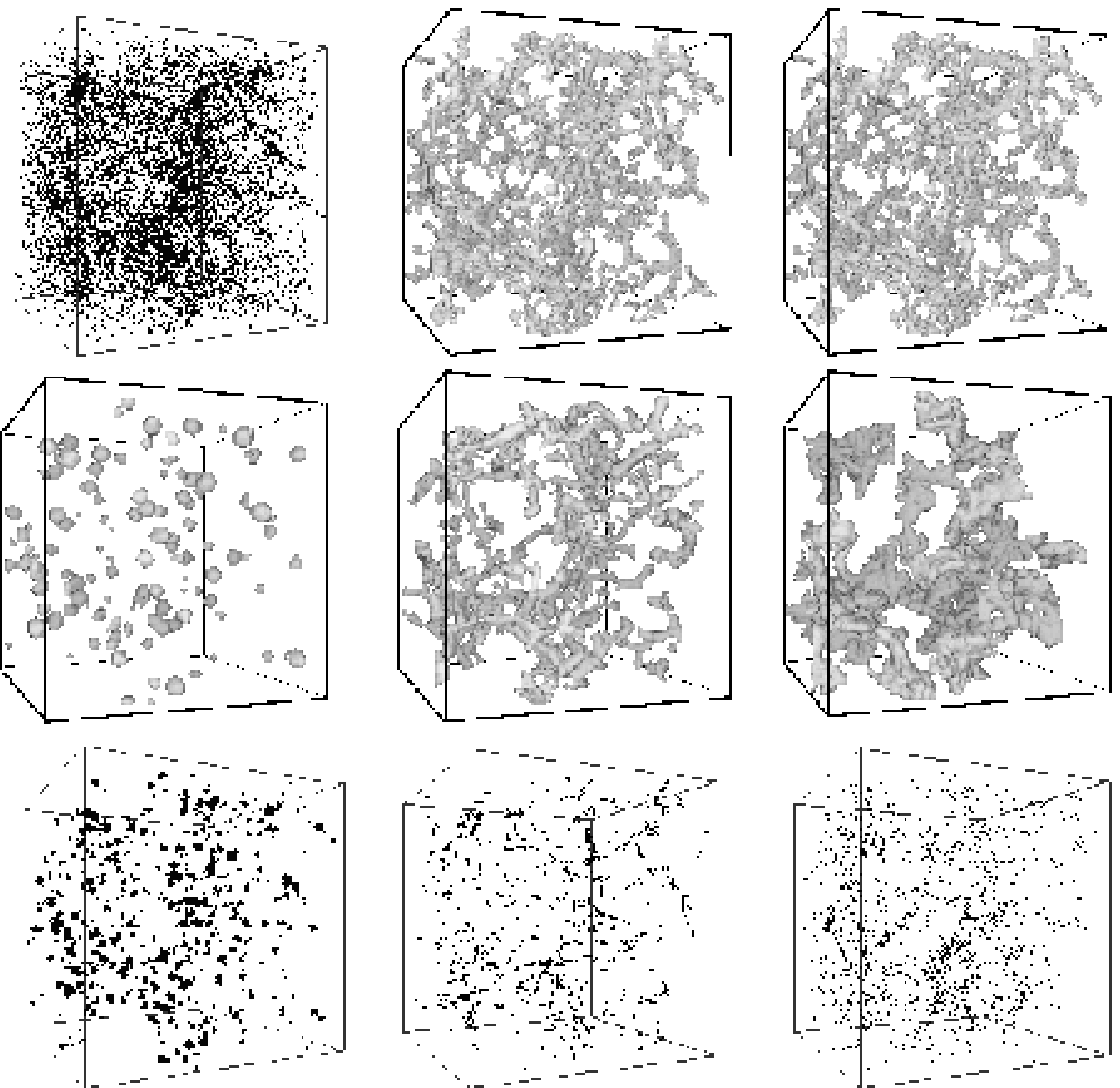}
    \caption{MMF applied to N-body simulation. The top row shows a subsample (a) consisting of 10 $\%$ of the total number of particles together with, in panels (b) and (c), the structures resulting from simple density thresholding using two different thresholds.  Panels (b) and (c) both contain both spherical and elongated structures: there is a large amount of cross contamination between morphologies.  Simple density thresholding is not an effective morphological discriminator.  The second row shows the results of applying the MMF procedure showing clearly segregated (a) blobs, (b) filaments and (c) walls (for clarity we display only the largest structures.  The third row shows the particles associated with the MMF defined structures.
}
  \label{fig:N_body_ALL}
\end{figure*}

In summary, we find that MMF clearly performs much better in tracing blob, filament and
wall galaxies than a pure threshold criterion would allow. By comparing Voronoi models
A, B, C and D we find that MMF performs better for components which are relatively more
prominent. Because of the mixture in densities between blobs, filaments and walls this is
not necessarily true when using a simple density criterion. The latter involves
often excessive levels of contamination between galaxies in different morphological
entities. If anything, this is perhaps the strongest argument for the use of the shape
and morphology criteria enclosed in the MMF.

\begin{figure*}[bhtp]
  \centering
     \mbox{\hskip -0.0truecm\includegraphics[width=0.9\textwidth]{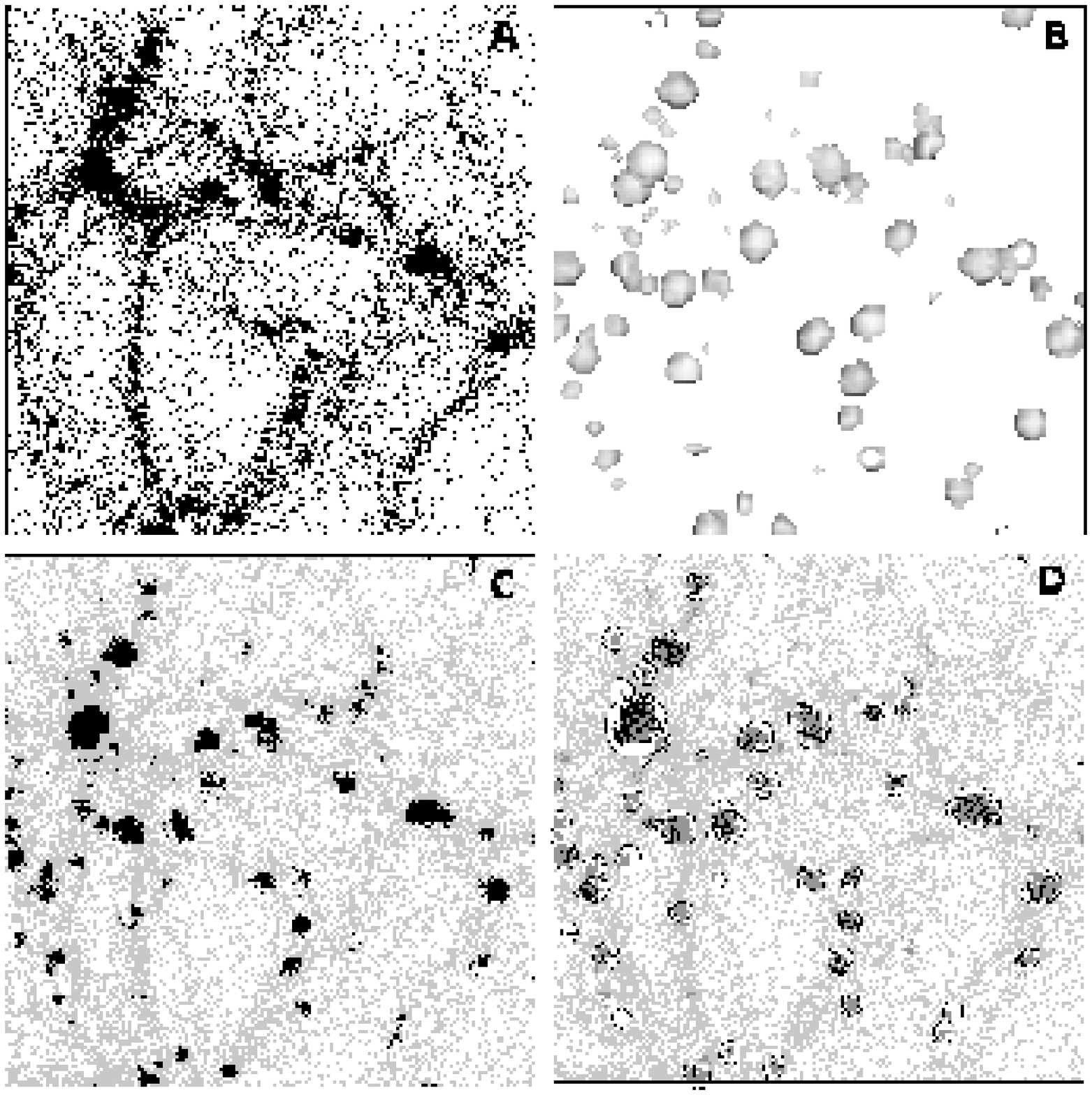}}
    \caption{Comparing blobs found from HOP and from MMF.    (A): Particles. (B): Isosurfaces of the blob identiffied with MMF.  (C): Particles inside the blobs (black) and background particles in grey. (D): The position of the HOP Haloes (circles) and the particles inside the MMF blobs (dark grey). Light grey particles are just the rest.
 }
  \label{fig:mmf_hop_blobs} 
\end{figure*}

%
%
\section{N-body simulations}
\label{sec:nbody}
The Large Scale Structure of the universe contains an intrincate mixture of morphologies.  The boundaries separating each morphological component is rather ill-defined: clusters of galaxies are interconnected by filaments which in turn define the edges of walls. 

In order to explore the response of the MMF in this complex scenario we performed a cosmological N-body simulation.  We give here only a few preliminary results to illustrate how the methodology works with a ``real'' galaxy distribution.  A more detailed exploratin follows in a later paper.
 
The simulation represents a LCDM model with $\Omega_{\Lambda}=0.7$, $\Omega_m=0.3$, $h=0.73$ in a periodic box of side 150Mpc containing $256^3$ dark matter particles.  We also run the same simulation lowering the resolution to $128^3$ particles according to the prescription given by \citet{klypin2001} in order to assess the effect of mass resolution in the density field determination. For the scales studied here there is no significant difference between the density fields computed from the two simulations since the mean interparticle separation is small enough to resolve the intermedium-density structures \citep{willemphd2007}.

\subsection{Results}
Figure \ref{fig:N_body_ALL} shows the result of applying the MMF to this simulation. The multiscale nature of the MMF is clearly seen in figure \ref{fig:N_body_ALL}c which shows blobs of different sizes containing similar sized clusters of points (Figure\ref{fig:N_body_ALL}g).

In the case of filaments and walls (see Figure\ref{fig:N_body_ALL}e and \ref{fig:N_body_ALL}f)  the multiscale nature of the MMF is less so obvious, however it is nonetheless there. 
 
It is clear from figure \ref{fig:N_body_ALL} that even though the LSS presents a great challenge it can succesfully recover each morphological component at its characteristic scale.

\begin{figure*}[htp]
  \centering
     \mbox{\hskip -0.0truecm\includegraphics[width=0.9\textwidth]{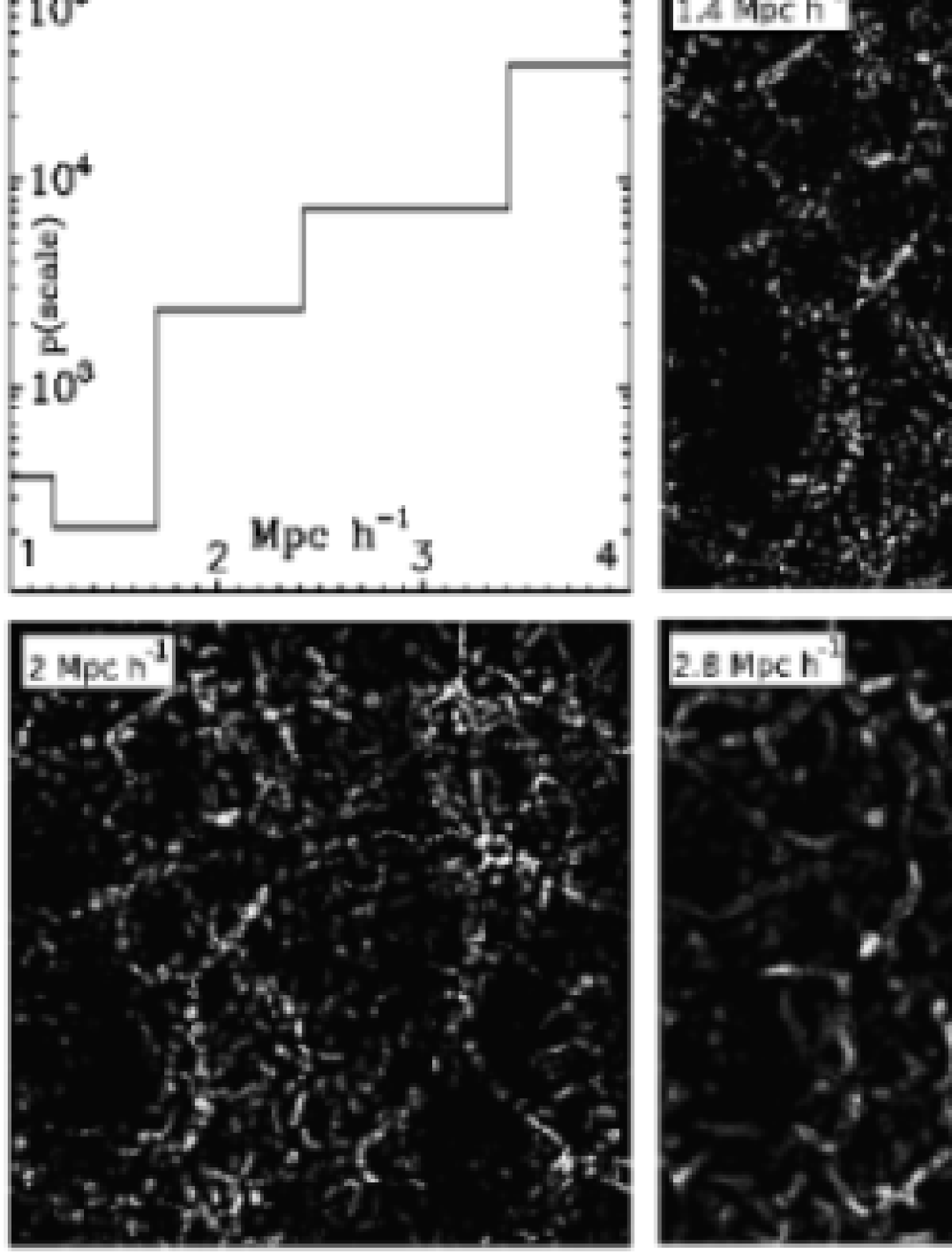}}
    \caption{Particles defining filamentary structures in a slice of an N-body model. The grayscale images show the 
MMF detection of filamentary features on various filtering scales. Top lefthand: the filament volume occupancy  
(number of sample grid cells with a filament signal) as a function of smoothing scale.
 }
  \label{fig:mmf_filaments} 
\end{figure*}

\subsection{Blobs and Clusters}
Figure~\ref{fig:mmf_hop_blobs}A shows a section through the N-body model and figure~\ref{fig:mmf_hop_blobs}B shows the blobs that are found in that section by the MMF process.  Figure~\ref{fig:mmf_hop_blobs}C shows how these blobs relate to the underlying structure displayed in panel A.

In an attempt to see how these blobs compare with clusters that would be identified via a more traditional method, we have use the HOP algorithm to find clusters.  The HOP clusters are shown in figure~\ref{fig:mmf_hop_blobs}D superposed on the MMF blobs.  Superficially, we see that the agreement is remarkably good: the MMF blobs are indeed what we would subjectively call clusters.  As in figure \ref{fig:N_body_ALL} we can appreciate the scale range revealed by MMF.

Making this comparison more precise is rather difficult owing the the vastly different approaches to blob-finding.  This will be discussed in detail in a subsequent paper dealing specifically with the application of MMF to N-body simulations.

\subsubsection{Filaments}
In Figure \ref{fig:mmf_filaments} we show the particles that belong to the filaments defined at various scales of the scale space.  The top left panel of figure~\ref{fig:mmf_filaments} shows a histogram of the number of particles contained in the filaments seen at smoothing scales from $1-4 h^{-1}$ Mpc.  As expected the number of particles rises rapidly with smoothing scale (the filaments are fatter on larger scales and so encompass greater volume).  The other three panels show the points contained in filaments, seen in a slice of the N-body simulation at different resolutions.  

When these are stacked, application of equation (\ref{eq:max_scale_value}) determines whether a given pixel is a part of a filament.  The process yields the filamentary map of figure~{\ref{fig:N_body_ALL}}
 
\subsection{Inventory of structures}
Finally we can simply count up the mass fraction of the model in various structural entities and the volume occupied by such structures to see how much of this N-body Universe lies in which structures.  The result is shown in the pie diagrams of Figure~\ref{fig:mmf_filaments}.

\begin{figure}[htp]
 \begin{center}
  \subfigure[Volume occupied by each of the structural features.]
	{
    \label{fig:mmf_pie_volume} 
  \includegraphics[width=5.0cm,angle=0.0]{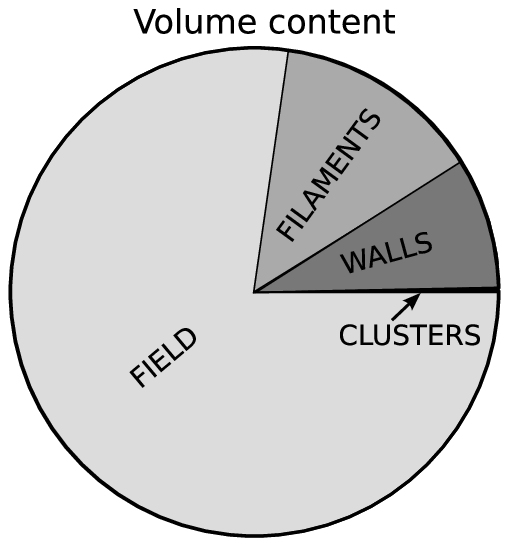}
	}
  \subfigure[Fraction of the mass occupancing each of the structural features.]
	{
    \label{fig:mmf_pie_mass} 
  \includegraphics[width=5.0cm,angle=0.0]{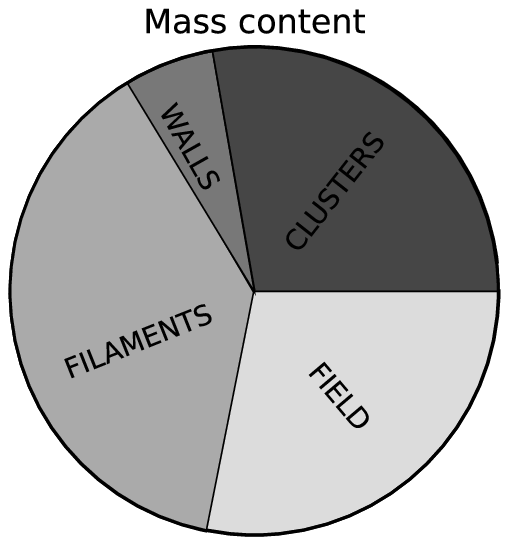}
	}
 \end{center}
\caption{Occupancy of Cosmic Web features, by volume (top) and mass(bottom) for a $\Lambda$CDM N-body simulation 
(see text).}
\label{fig:Pies}
\end{figure}

The result is hardly surprising: the clusters and filaments occupy about the same mass fraction and together contain more than half the haloes in the simulation.  The clusters occupy by far least volume - they are dense systems and they are denser than the filaments.  Recall, however, the important remark that we could not use density threshold alone to define these structures (see the top row of panels in figure \ref{fig:N_body_ALL}).
  
The large volume occupancy of filamentary structures explains why our impression of the cosmic matter distribution is predominantly filamentary, and the fact that they are all long and thin (as illustrated in figure \ref{fig:mmf_filaments}) 
emphasises the web-like nature of the structure.  

Perhaps the only surprise in this analysis is the relatively low volume occupancy of the walls in comparison with the filaments.  This may be in part because most of the walls have broken up by the present epoch.  It may also be in part due to the fact that the low number of particles in walls makes it relatively difficult to find them: they may get mis-classified as being part of the general field.  It is difficult to assess this on the basis of the present experiments alone.

%
%
\section{Conclusions and comments}
MMF, our simple embodiment of Hessian-based scale space analysis of large scale cosmic structure, is remarkably sucessful in delineating the different structures that make up the cosmic web.   Since the morphology filters give us a direct measurement of blobness, filamentariness or wallness they can be used to characterize and quantify, in a systematic way, the large scale matter distribution. The technique has been tested using N-Body and Voronoi models.
 
\subsection{Void finding}
It should be emphaised that MMF is not a void finder except insofar as anything that is not in a blob, filament of wall might be deemed to be in a void region.  In that case MMF would be a suitable tool for finding so-called ``void galaxies'' without being able to identify the host void.  Void finding \textit{per se} is almost certainly best achieved via the Watershed (WVF) method \cite{platen2007}.

\subsection{Enhancements} 
There are many areas where the MMF treatment could be enhanced and some of these will be presented in future papers.  We summarize a few issues here in order to place the present work in a more general perspective.

\begin{itemize}
\item[$\bullet$] The definition of the intensity component of the morphology filter could be improved by including other local propeties such as gradient, direction of eigenvectors, connectivity, etc.   
\item[$\bullet$] The Gaussian kernel is not the only possibility for producing the scale-space representation: alternative kernels may improve the performance of the MMF.  One side effect of using a simple Gaussian filter is that high peaks in high density filaments are detected always at larger scales even when their density profile is relatively narrow (the ``filter smearing'' we referred to earlier).
\item[$\bullet$] Our implementation of the Multiscale Morphology Filters is grid-based and that required a resampling of the original point distribution data.  It is possible to derive a similar set of filters using particle-based measures for the local distribution of matter (e.g.: inertia tensor analysis), defining window functions and scale normalizations in a multiscale context.
\end{itemize}
With respect to the above we would also like to refer to sect.~\ref{sec:developments}.
 
\subsection{Applications}
The ability to accurately identify arbitrarily shaped structures allows the possibility of seeking correlations within the structures that might otherwise be masked by other methods.  Already, the method has been used to identify previoulsy unknown systemic properties in the alignment of haloes with the parent structures \citep{aragon2007}.

The technique has been illustrated in terms of spatial point data since that is relatively unambiguous.  However, the MMF technique we have described is a quite general technique for scale-free feature finding: it only needs a mathematical prescription of what is being looked for, which in general may not be so easy!  Bearing that in mind, the following is a list of possible application areas.

The technique can readily be extended to analysis of velocity data of various kinds such as Fingers Of God in cosmological redshift surveys, analysis of dynamical phase spaces, feature detection in solar images, morphological characterization of structure in spiral arms, feature detection in radio datacubes, etc.  Finding clusters and their substructures using MMF would provide an important alternative to HOP.  Finding small, low surface brightness, galaxies in noisy neutral hydrogen surveys would be another useful application.


\begin{acknowledgements}
We would like to thank Pablo Araya for providing the N-body simulations and 
Erwin Platen for many useful and clarifying discussions. 
\end{acknowledgements}
 

\appendix
 

%
%
\section{\ \\ The DTFE general reconstruction procedure}
\label{app:dtfe_recons}
For a detailed specification of the DTFE density field procedure we refer to \cite{willemphd2007}. 
In summary, the DTFE procedure for density field reconstruction from a discrete set of points consists of the
following steps:
\begin{enumerate}
\item[$\bullet$] {\bf Point sample}\\ Given that the point sample is
supposed to represent an unbiased reflection of the underlying density
field, it needs to be a general Poisson process of the (supposed)
underlying density field.
\medskip
\item[$\bullet$] {\bf Boundary Conditions}\\ The boundary conditions
will determine the Delaunay and Voronoi cells that overlap the
boundary of the sample volume. Dependent on the sample at hand, a
variety of options exists:
\begin{enumerate}
\item[+] {\it Empty boundary conditions:}\\ outside the sample volume
there are no points.
\item[+] {\it Periodic boundary conditions:}\\ the point sample is
supposed to be repeated periodically in boundary boxes, defining a
toroidal topology for the sample volume. 
\item[+] {\it Buffered boundary conditions:}\\ the sample volume box is
surrounded by a bufferzone filled with a synthetic point sample.
\end{enumerate}
\medskip
\item[$\bullet$] {\bf Delaunay Tessellation}\\ Construction of the
Delaunay tessellation from the point sample.  While we also still use the
Voronoi-Delaunay code of \cite{weygaertphd1991} and \cite{weygaert1994}, at present 
there is a number of efficient library routines available. Particularly noteworthy
is the \cgal initiative, a large library of computational geometry
routines\footnote{\cgal is a \texttt{C++} library of algorithms and
data structures for Computational Geometry, see \url{www.cgal.org}.}\\
\medskip
\item[$\bullet$] {\bf Field values point sample}\\ The estimate of the
density at each sample point is the normalized inverse of the volume
of its contiguous Voronoi cell ${\cal W}_i$ of each point
$i$. The {\it contiguous Voronoi cell} of a point $i$ is the union of
all Delaunay tetrahedra of which point $i$ forms one of the four
vertices. We recognize two applicable situations:
\\ \itemitem{-} {\it uniform sampling process}:\\ 
the point sample is an unbiased sample of
the underlying density field. Typical example is that of $N$-body
simulation particles. For $D$-dimensional space the density estimate
is,
\begin{equation}
{\widehat \rho}({\bf x}_i)\,=\,(1+D)\,\frac{w_i}{V({\cal W}_i)} \,.
\label{eq:densvor}
\end{equation}
\noindent with $w_i$ the weight of sample point $i$, usually we assume
the same ``mass'' for each point. \\ 
\itemitem{-} {\it systematic non-uniform sampling process}:\\
 sampling density according to specified
selection process.  The non-uniform sampling process is quantified by
an a priori known selection function $\psi({\bf x})$. This situation
is typical for galaxy surveys, $\psi({\bf x})$ may encapsulate
differences in sampling density $\psi(\alpha,\delta)$ as function of
sky position $(\alpha,\delta)$, as well as the radial redshift
selection function $\psi(r)$ for magnitude- or flux-limited
surveys. For $D$-dimensional space the density estimate is ,
\begin{equation}
{\widehat \rho}({\bf x}_i)\,=\,(1+D)\,\frac{w_i}{\psi({\bf x}_i)\,V({\cal W}_i)} \,.
\label{eq:densvornu}
\end{equation}

\medskip
\item[$\bullet$] {\bf Field Gradient}\\ Calculation of the field
gradient estimate $\widehat{\nabla f}|_m$ in each $D$-dimensional
Delaunay simplex $m$ ($D=3$: tetrahedron; $D=2$: triangle) by solving
the set of linear equations for the field values $f_i$ at the positions 
${\bf r}_i$ of the $(D+1)$ tetrahedron vertices,\\
\begin{eqnarray}
\widehat{\nabla f}|_m \ \ \Longleftarrow\ \ 
\begin{cases}
f_0 \ \ \ \ f_1 \ \ \ \ f_2 \ \ \ \ f_3 \\
\ \\
{\bf r}_0 \ \ \ \ {\bf r}_1 \ \ \ \ {\bf r}_2 \ \ \ \ {\bf r}_3 \\
\end{cases}\,
\label{eq:dtfegrad}
\end{eqnarray}
Evidently, linear interpolation for a field $f$ is only meaningful
when the field does not fluctuate strongly.

\medskip
\item[$\bullet$] {\bf Interpolation}.\\ The final basic step of the
DTFE procedure is the field interpolation. The processing and
postprocessing steps involve numerous interpolation calculations, for
each of the involved locations ${\bf x}$. Given a location ${\bf x}$, the 
Delaunay tetrahedron $m$ in which it
is embedded is determined. On the basis of the field gradient
$\widehat{\nabla f}|_m$ the field value is computed by (linear)
interpolation,
\begin{equation}
{\widehat f}({\bf x})\,=\,{\widehat f}({\bf x}_{i})\,+\,{\widehat {\nabla f}} \bigl|_m \,\cdot\,({\bf x}-{\bf x}_{i}) \,.
\label{eq:fieldval}
\end{equation}
In principle, higher-order interpolation procedures are also
possible. Two relevant procedures are:\\ 
\item[-] {\it Spline Interpolation} 
\item[-] {\it Natural Neighbour Interpolation}\\ 
\item[]
\noindent For NN-interpolation see \cite{watson1992,braunsambridge1995,sukumarphd1998} and 
\cite{okabe2000}. Implementation of Natural neighbour interpolations is presently 
in progress.
\medskip
\item[$\bullet$] {\bf Processing}.\\ Though basically of the same
character, for practical purposes we make a distinction between
straightforward processing steps concerning the production of images
and simple smoothing filtering operations and more
complex postprocessing.  The latter are treated in the next item. 
Basic to the processing steps is the determination of
field values following the interpolation procedure(s) outlined
above. Straightforward ``first line'' field operations are {\it
Image reconstruction} and {\it Smoothing/Filtering}.

\medskip
\begin{enumerate}
\item[+] {\bf Image reconstruction}.\\ For a set of image
points, usually grid points, determine the image value. 
formally the average field value within the corresponding gridcell.
In practice a few different strategies may be followed\\
\item[-] {\it Formal geometric approach} 
\item[-] {\it Monte Carlo approach}
\item[-] {\it Singular interpolation approach}\\
\item[]
\noindent The choice of strategy is mainly dictated by accuracy requirements. For WVF we 
use the Monte Carlo approach in which the grid density value is the average 
of the DTFE field values at a number of randomly sampled points within the grid 
cell. \\
\item[+] {\bf Smoothing} and {\bf Filtering}:\\
A range of filtering operations is conceivable. Two of relevance to WVF are:\\
\item[-] {\it Linear filtering} of the field ${\widehat f}$\\ 
Convolution of the field ${\widehat f}$ with a filter \\
function $W_s({\bf x},{\bf y})$, usually user-specified, 
   \begin{equation}
     f_s({\bf x})\,=\,\int\,{\widehat f}({\bf x'})\, W_s({\bf x'},{\bf y})\,d{\bf x'}     
   \end{equation}
\item[-] {\it Natural Neighbour Rank-Ordered filtering}\\ 
\citep{platen2007}.  
\item[]
\end{enumerate}
\item[$\bullet$] {\bf Post-processing}.\\ The real potential of DTFE
fields may be found in sophisticated applications, tuned towards
uncovering characteristics of the reconstructed fields.  An important
aspect of this involves the analysis of structures in the density
field. The WVF formalism developed in this study is an obvious
example.
\end{enumerate}

%
\begin{figure*}
  \centering
     \includegraphics[width=0.95\linewidth]{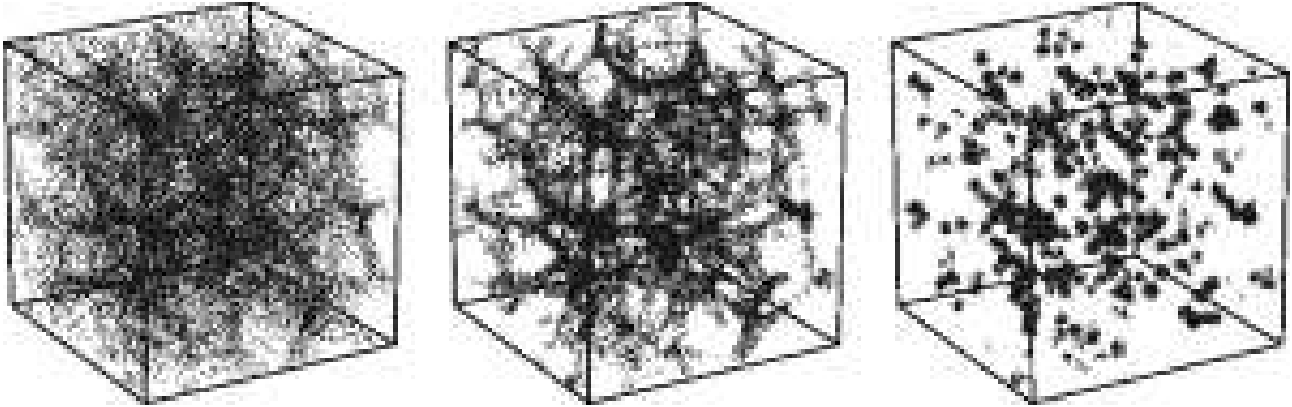}
     \caption{\small Three different patterns of Voronoi element galaxy distributions, 
           shown in a 3-D cubic setting. The depicted spatial distributions 
           correspond to a wall-dominated Voronoi Universe (top), a filamentary 
           Voronoi Universe (centre) and a cluster-dominated Voronoi Universe 
           (bottom). }  
\label{fig:vorelm3}
\end{figure*}

%
%
\section{\ \\Voronoi clustering models}
\label{app:vorclustform}
{\it Voronoi Clustering Models} are a class of heuristic models for cellular distributions of matter  
\citep{weygaertphd1991,weygaert2007}. They use the Voronoi tessellation as the skeleton of the cosmic 
matter distribution, identifying the structural frame around which matter will gradually assemble during 
the emergence of cosmic structure. The interior of Voronoi {\it cells} correspond to voids and the Voronoi {\it planes} 
with sheets of galaxies. The {\it edges} delineating the rim of each wall are identified with the filaments in the galaxy distribution. 
What is usually denoted as a flattened ``supercluster'' will comprise an assembly of various connecting walls in the 
Voronoi foam, as elongated ``superclusters'' of ``filaments'' will usually consist of a few coupled edges. The most 
outstanding structural elements are the {\it vertices}, corresponding to the very dense compact nodes within the 
cosmic web, rich clusters of galaxies. 
 
We distinguish two different yet complementary approaches \citep[see][]{weygaert2007}. One is the fully heuristic approach 
of {\it ``Voronoi Element models''}. They are particularly apt for studying systematic properties of spatial galaxy distributions 
confined to one or more structural elements of nontrivial geometric spatial patterns. The second, supplementary, approach is that 
of the {\it Voronoi Evolution models} or {\it Voronoi Kinematic models}, which attempt to ``simulate'' weblike galaxy distributions 
on the basis of simplified models of the evolution of the Megaparsec scale distribution. 
The Voronoi clustering models offer flexible templates for cellular patterns, and they are easy to tune towards a particular 
spatial cellular morphology. To investigate the performance of MMF we use composite {\it Voronoi Element Models}, tailor-made 
heuristic ``galaxy'' distributions composed of a superposition of particle distributions in and around the walls, edges and 
vertices of the Voronoi skeleton. A complete composite particle distribution includes particles located in four 
distinct structural components:
\begin{enumerate}
\item[$\bullet$] \emph{Field}\\ Particles located in the {\it interior of Voronoi cells} (and thus randomly distributed across 
the entire model box)  
\item[$\bullet$] \emph{Wall}\\ Particles within and around the {\it Voronoi walls}.
\item[$\bullet$] \emph{Filament}\\Particles within and around the {\it Voronoi edges}. 
\item[$\bullet$] \emph{Blobs}\\ Particles within and around the {\it Voronoi vertices}. 
\end{enumerate}
For each of these four distinct distributions the model galaxies are projected onto the relevant Voronoi wall, Voronoi edge 
or Voronoi vertex or retained within the interior of the Voronoi cell in which they are located. Characteristic examples of 
{\it simple Voronoi Element} galaxy distributions are the ones shown in the boxes in Fig.~\ref{fig:vorelm3}. The 
depicted distributions concern a wall-dominated Voronoi Universe (lefthand), a filamentary Voronoi Universe (centre) and 
a cluster-dominated Voronoi Universe (righthand). 

In the case of composite models the fraction of {\it field galaxies} $X_c$, {\it wall galaxies} $X_w$, {\it filaments galaxies} $X_f$ 
and {\it blob galaxies} $X_b$, with the constraint $X_c+X_w+X_f+X_b=100$, is a key input parameter of the model. 

\subsection{Initial Conditions}
\label{app:vorclustform}
The initial conditions for the Voronoi galaxy distribution are:
\begin{itemize}
\item[$\bullet$] Distribution of $M$ nuclei, {\it expansion centres}, within the simulation volume $V$. The location 
of nucleus $m$ is ${\bf y}_m$.
\item[$\bullet$] Generate $N$ model galaxies whose initial locations, ${\bf x}_{n0}$ $(n=1,\ldots,N)$, are randomly 
distributed throughout the sample volume $V$. 
\item[$\bullet$] Of each model galaxy $n$ determine the Voronoi cell ${\cal V}_{\alpha}$ in which it is located, ie. 
determine the closest nucleus $j_{\alpha}$.
\end{itemize}

\noindent All different Voronoi models are based upon the displacement
of a sample of $N$ ``model galaxies''. The initial spatial
distribution of these $N$ galaxies within the sample volume $V$ is
purely random, their initial locations ${\bf x}_{n0}$ ($n=1,\ldots,N)$
defined by a homogeneous Poisson process. A set of $M$ nuclei within the volume 
$V$ corresponds to the cell centres, or {\it expansion centres} driving the 
evolving matter distribution. The nuclei have locations ${\bf y}_m$ $(m=1,\ldots,M)$.

Following the specification of the initial positions of all galaxies,
the second stage of the procedure consists of the calculation of the
complete Voronoi track for each galaxy $n=1,\ldots,N$ (sec.~\ref{sec:vortrack}) 
towards its wall, filament or vertex, or its location within a cell when 
it is a field galaxy. . 

{\it Simple Voronoi Element Models} place all model galaxies in either walls, edges or vertices. The 
versatility of the model also allows combinations of element models, in which field (cell), wall, 
filament and vertex distributions are superimposed. The characteristics of the patterns and 
spatial distribution in these {\it Mixed Voronoi Element Models} can be varied and tuned according to the 
fractions of wall galaxies, filament galaxies, vertex and field galaxies. 

\begin{figure}[t]
  \centering
  \vskip 0.5truecm
  \mbox{\hskip 0.0truecm\includegraphics[height=7.0cm]{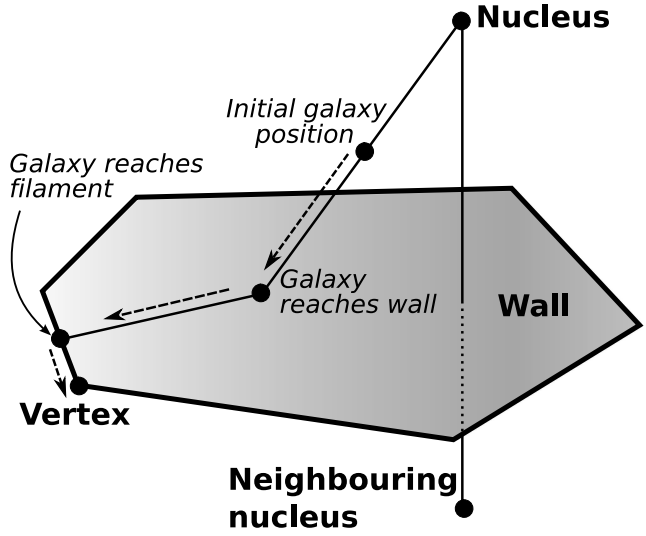}}
  \vskip 0.2cm
  \caption{Schematic illustration of the galaxy projections in the Voronoi clustering model. See text.}
  \label{fig:vorkinmschm}
\end{figure} 
\subsection{Voronoi Tracks}
\label{sec:vortrack}
The first step of the formalism is the determination for each galaxy $n$ the Voronoi cell ${\cal V}_{\alpha}$ in which it 
is initially located, ie. finding the nucleus $j_{\alpha}$ which is closest to the galaxies' initial position ${\bf x}_{n0}$. 

In the second step the galaxy $n$ is moved from its initial position ${\bf x}_{n0}$, towards its final 
destination in a wall, filament or vertex (see fig.~\ref{fig:vorkinmschm}). The first section of the galaxy displacement is 
the radial path along the direction defined by the galaxies' initial location wrt. its expansion centre $j_\alpha$. This 
direction is defined by the unity vector ${\hat {\bf e}}_{n\alpha}$.
 
If the galaxy is a field galaxy it remains at its original location. If it is a wall galaxy it is 
projected along direction ${\hat {\bf e}}_{n\alpha}$ onto the Voronoi wall $\Sigma_{\alpha\beta}$ with 
which the radial path first intersects. Filament galaxies are moved along the wall to the location where 
the path intersects $\Lambda_{\alpha\beta\gamma}$. Finally, if it is cluster galaxy the galaxies' path 
is continued along the edge $\Lambda_{\alpha\beta\gamma}$ until it reaches its final destination, 
vertex $\Xi_{\alpha\beta\gamma\delta}$. The identity of the neighbouring nuclei $j_{\alpha}$, $j_{\beta}$, 
$j_{\gamma}$ and $j_{\delta}$, and therefore the identity of the cell ${\cal V}_{\alpha}$, the wall $\Sigma_{\alpha\beta}$, 
the edge $\Lambda_{\alpha\beta\gamma}$ and the vertex $\Xi_{\alpha\beta\gamma\delta}$, depends on the initial location 
${\bf x}_{n0}$ of the galaxy, the position ${\bf y}_{\alpha}$ of its closest nucleus and the definition of the galaxies' path 
within the Voronoi skeleton. 

\bigskip
\noindent In summary, the path ${\b x}_n$ is codified by 
\begin{eqnarray}
{\bf x}_n&\,=\,&{\bf y}_{\alpha}\,+\,{\bf s}_{n\alpha}\,+\,{\bf s}_{n\alpha \beta}\,+\,{\bf s}_{n\alpha \beta \gamma} \nonumber\\
\ \\
&\,=\,&{\bf y}_{\alpha}\,+\,s_{n\alpha}{\hat {\bf e}}_{n\alpha}\,+\,s_{n\alpha\beta}{\hat {\bf e}}_{n\alpha\beta}\,+\,s_{n\alpha\beta\gamma}
{\hat {\bf e}}_{n\alpha\beta\gamma}\nonumber
\label{eq:galpath}
\end{eqnarray}
\noindent in which the four different components follow the directions defined by:
\begin{enumerate}
\item[$\bullet$] ${\hat {\bf e}}_{n\alpha}$: \\ unity vector of path within Voronoi cell ${\cal V}_{\alpha}$
\item[$\bullet$] ${\hat {\bf e}}_{n\alpha\beta}$:\\unity vector of path within Voronoi wall $\Sigma_{\alpha\beta}$
\item[$\bullet$] ${\hat {\bf e}}_{n\alpha\beta\gamma}$:\\ unity vector of path along Voronoi edge $\Lambda_{\alpha\beta\gamma}$ 
\item[$\bullet$] Vertex $\Xi_{\alpha\beta\gamma\delta}$
\end{enumerate}

The cosmic matter distribution is obtained by calculating the individual displacement factors $(s_{n\alpha},s_{n\alpha\beta},
s_{n\alpha\beta\gamma})$ for each model galaxy, corresponding to their location within either wall, filament or vertex. In the 
Voronoi Element models all galaxies are directly projected onto wall, edge or vertex following the path depicted in 
fig.~\ref{fig:vorkinmschm}. The corresponding displacement factors in eqn.~\ref{eq:galpath} for a wall, filament or cluster 
galaxy are
\begin{eqnarray}
{\rm Walls}\quad(s_{n\alpha},s_{n\alpha\beta},s_{n\alpha\beta\gamma})&\,=\,&(\upsilon_n,0,0)\nonumber\\
\ \nonumber\\
{\rm Filaments}\quad(s_{n\alpha},s_{n\alpha\beta},s_{n\alpha\beta\gamma})&\,=\,&(\upsilon_n,\sigma_n,0)\\
\ \nonumber\\
{\rm Clusters}\quad(s_{n\alpha},s_{n\alpha\beta},s_{n\alpha\beta\gamma})&\,=\,&(\upsilon_n,\sigma_n,\lambda_n)\nonumber
\label{eq:vorelmfac}
\end{eqnarray}
where the values of the parameters $\upsilon_n$, $\sigma_n$ and $\lambda_n$ characterize the crossing of the 
galaxies' path with the wall, edge or vertex towards which it moves. 

A finite thickness is assigned to all Voronoi structural elements. The walls, filaments and vertices are assumed to have a 
Gaussian radial density distribution specified by the widths $R_{\rm W}$ of the walls, $R_{\rm F}$ of the filaments 
and $R_{\rm V}$ of the vertices. Voronoi wall galaxies are displaced according to the specified Gaussian density profile in
the direction perpendicular to their wall. A similar procedure is followed for the Voronoi filament galaxies and
the Voronoi vertex galaxies. As a result the vertices stand out as three-dimensional Gaussian peaks.


\end{document}